\documentclass[twocolumn,iop,fleqn]{openjournal}

\usepackage[english]{babel}
\usepackage{graphicx}	
\usepackage{amsmath}	
\usepackage{amssymb}	
\usepackage{textgreek}  
\usepackage{longtable}  
\usepackage{tocbibind}
\usepackage[toc,page]{appendix}
\usepackage{newtxtext,newtxmath}
\usepackage[T1]{fontenc}
\usepackage{amsmath}
\usepackage{graphicx}
\usepackage{natbib}
\usepackage{comment}
\usepackage{hyperref}
\extrafloats{110}
\usepackage[maxfloats=110]{morefloats}
\usepackage{array}
\usepackage{tabularx}   

\begin{document}

\title{Near-Infrared Spectroscopy of Low-Redshift Palomar-Green Quasars \\
and Composite Spectra}

\author{Amy Cavanaugh$^{1}$}
\author{Michael Brotherton$^{1}$}
\author{Jacob McLane$^{1}$}
\author{Richard Green$^{2}$}

\affiliation{$^{1}$ Department of Physics and Astronomy, University of Wyoming
1000 E University Ave, Laramie, WY 82071, USA}

\affiliation{$^{2}$Steward Observatory, University of Arizona, 933 North Cherry Avenue, Tucson, AZ 85721, USA}

\begin{abstract}

The Palomar-Green (PG) Quasars have been used extensively to establish
the spectral properties of luminous active galactic nuclei (AGNs) at low redshifts
in many wavebands, but not the near-infrared (NIR).  We present NIR spectra 
of all 87 PG quasars with $z < 0.5$ observed with the Apache Point Observatory 3.5-m telescope and the TripleSpec spectrograph in standard cross-dispersed mode.  Here we employ our data set to construct a NIR composite quasar
spectrum spanning rest-frame wavelengths from approximately 8800-22000 \AA\, reaching a signal-to-noise ratio of several hundred in the H-band for 4 \AA\ pixels.
We characterize the continuum and emission-line properties of this 
composite.  Additionally, we construct composite spectra as a function
of luminosity and Eddington ratio, finding some differences, not entirely unexpected.  In particular, more luminous composite NIR spectra grow increasingly red at long wavelengths, consistent with relatively stronger dust 
emission at higher temperatures.

\end{abstract}


\section{Introduction} \label{sec:introduction}

Quasars emit light across the entire electromagnetic spectrum, with each waveband offering features that arise from one or more physically distinct regions.  Carefully selected samples of quasars can be observed to establish the spectral properties of different wavebands, which depend on the underlying physical structures and their associated physics.

One sample extensively investigated to determine the observable properties of quasars and the relationships among them is a subset of the Palomar-Green Survey (PG) \citep{1983ApJ...269..352S} called the Bright Quasar Survey (BQS) that contains stellar objects displaying an ultraviolet excess. Specifically, the BQS has an average limiting magnitude of $B = 16.16$ and an ultraviolet excess of $U -B < -0.44$. The 92 BQS quasars additionally show broad emission lines and comprise a significant fraction of the most luminous AGN at low redshift. Although PG quasars were originally advertised as a nearly complete sample, they are not, although they appear to be representative \citep{2000A&A...358...77W,jester2005}. In fact, \citet{jester2005} compare BQS data to SDSS data to simulate PG object selection, finding a bluer U-B cutoff of about $-$0.7. 
\citet{1992ApJS...80..109B} focused on the 87 PG quasars with $z < 0.5$,
which includes 70 radio-quiet objects and 17 radio-loud objects.  They obtained
optical spectra and measured their emission-line properties.
A principal component analysis of these shows that two eigenvectors account for much of the variance: EV1 or PC1 (eigenvector 1 or principal component 1) and EV2 or PC2 (eigenvector 2 or principal component 2).  EV1 is primarily driven by the anticorrelation between the strengths of broad optical Fe II and narrow [O III] \textlambda5007 emission.  EV2 is mainly driven by the anticorrelation between luminosity and He II \textlambda4686 equivalent width. Several follow-up studies have solidified these results
and established that EV1 is associated with the Eddington ratio, L$_{\rm Bol}$ / L$_{\rm Edd}$
\citep[e.g.,][]{sulentic2000,2002ApJ...565...78B,2014Natur.513..210S,2015ApJ...804L..15S}.

This and other subsamples of the BQS have been investigated to establish quasar
properties across the electromagnetic spectrum.  For instance:
the X-rays \citep{1997ApJ...477...93L,2005A&A...432...15P},
the UV \citep{1999ApJ...515L..53W, 2007AJ....134..294S},
the mid-IR \citep{2007ApJ...666..806N,2014ApJS..214...23S,bian2016},
the far-IR \citep{2014ApJS..214...23S,2015ApJS..219...22P},
and the radio \citep{1989AJ.....98.1195K}.
Many X-ray, UV, and radio properties are tightly correlated with optical EV1
trends; understanding the physical processes of how the accretion rate drives
these observables is ongoing.

The rest-frame near-infrared (NIR) spectral properties have been less well studied than many other wavebands, for PG quasars and AGN in general, although some small spectral atlases have been published 
\citep{2006A&A...457...61R,2018ApJS..238...37K}.
The near-infrared displays the continuum emission characteristic of a power-law extending from the optical \citep{2008Natur.454..492K} 
likely emitted by an accretion flow with a disk-like geometry, 
plus a component that increases toward longer wavelengths, likely emitted by hot dust near the sublimation temperature \citep[e.g.,][]{2011MNRAS.414..218L}.  The near-infrared also sports a number of broad emission lines, primarily from the Paschen series of hydrogen, and a few narrow emission lines, notably [S III] $\lambda$9531 \citep{2006A&A...457...61R}.

\citet{Glikman2006} produced an NIR composite quasar spectrum of moderate signal-to-noise ratio using a sample of 27 objects selected from the Sloan Digital Sky Survey.  Their selection criteria were K$_s <14.5$, M$_i<-23$, and  
$0.118<z<0.418$.  In composite spectra, common features stand out while unusual ones cancel out, and they often represent the observable properties of typical objects.  

We conducted a NIR spectral survey of the same 87 $z < 0.5$ PG quasars,
as studied in the optical by Boroson \& Green (1992).
Our first steps are to produce composite NIR spectra for the entire sample,
as well as for subsamples that differ in terms of luminosity and L$_{\rm Bol}$/L$_{\rm Edd}$.
In section 2 we discuss data collection and reduction, and in section 3 we discuss the methods for constructing our composite spectra.  This work also uses the luminosity values from \citet{2006ApJ...641..689V}, who use a cosmology of H$_{0}$ = 70 km s$^{-1}$ Mpc$^{-1}$, \textOmega$_{\Lambda}$ = 0.7, and \textOmega$_{m}$ = 0.3.

\section{Data}

\subsection{Sample Properties}
We adopted the rest-frame monochromatic luminosity at 5100 \AA\ and black hole masses for the PG quasars from \citet{2006ApJ...641..689V}.   Bolometric luminosities were found using the methods of \citet{2012MNRAS.426.2677R}. The Eddington luminosities were found using the formula of \citet{1999agnc.book.....K}.  The specific relations are:

\begin{equation}
    \mathrm{log}(M_{\mathrm{BH}}) = \mathrm{log}\Bigr[\frac{\mathrm{FWHM}(\mathrm{H}\beta)}{1000\ \mathrm{km\ s^{-1}}}\Bigr]^{2}\Bigr[\frac{\lambda L_{\lambda}(5100\ \mathrm{\AA})}{10^{44}\ \mathrm{erg\ s^{-1}}}\Bigr]^{0.5} + 6.91
\end{equation}

\begin{equation}
    \mathrm{log}(L_{\mathrm{bol}}) = 4.891 + (0.912)\mathrm{log}(\lambda L_{\lambda}(5100))
\end{equation}

\begin{equation}
    L_{\mathrm{Edd}} = 1.51 \times 10^{38} \frac{M} {M_{\odot}}\ \mathrm{erg\ s^{-1}}
\end{equation}

The Eddington ratio is simply the ratio of the bolometric luminosity to the Eddington luminosity. 

\begin{figure}[!ht]
\includegraphics[scale=0.57]{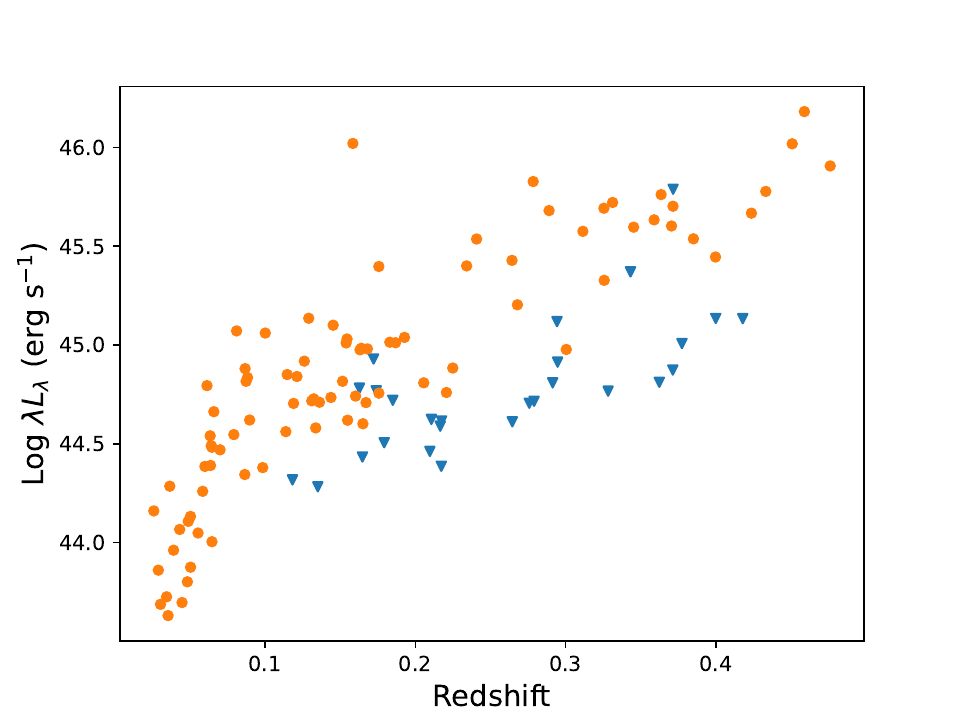}
\caption{\label{figl}Monochromatic luminosity at 5100 \AA\ for the 87 objects analyzed by \citet{1992ApJS...80..109B} (orange circles) and the SDSS-selected sample used by \citet{Glikman2006} (blue triangles).}

\end{figure}

\begin{figure}[!ht]
\centering
\includegraphics[scale=0.57]
{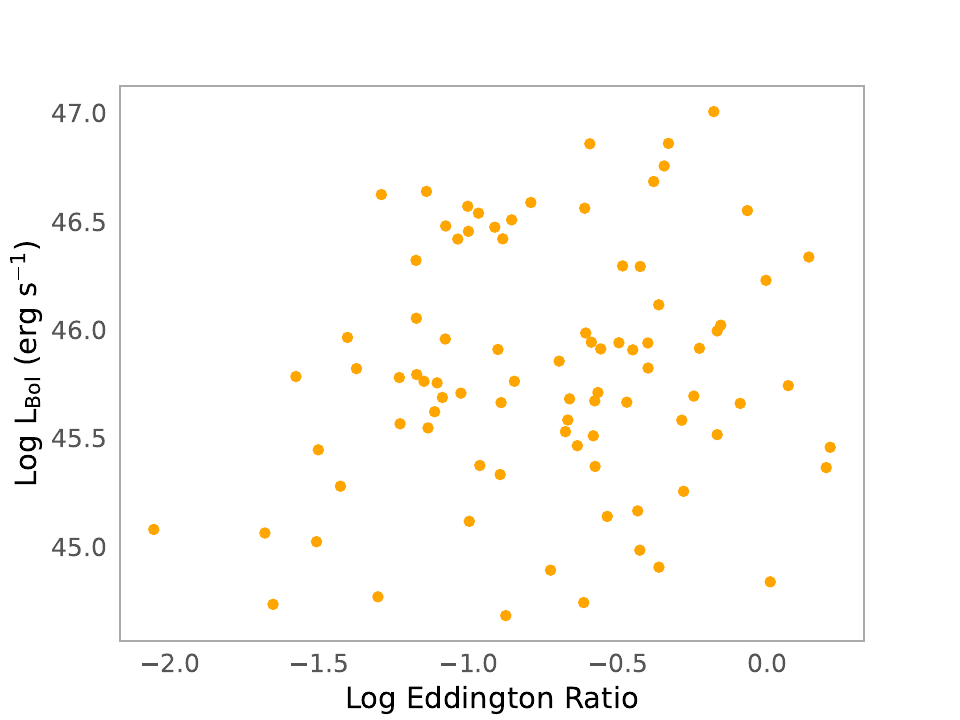}
\caption{\label{figboledd}Log bolometric luminosity and Eddington ratio distribution.  
}

\centering
\includegraphics[scale=0.57]
{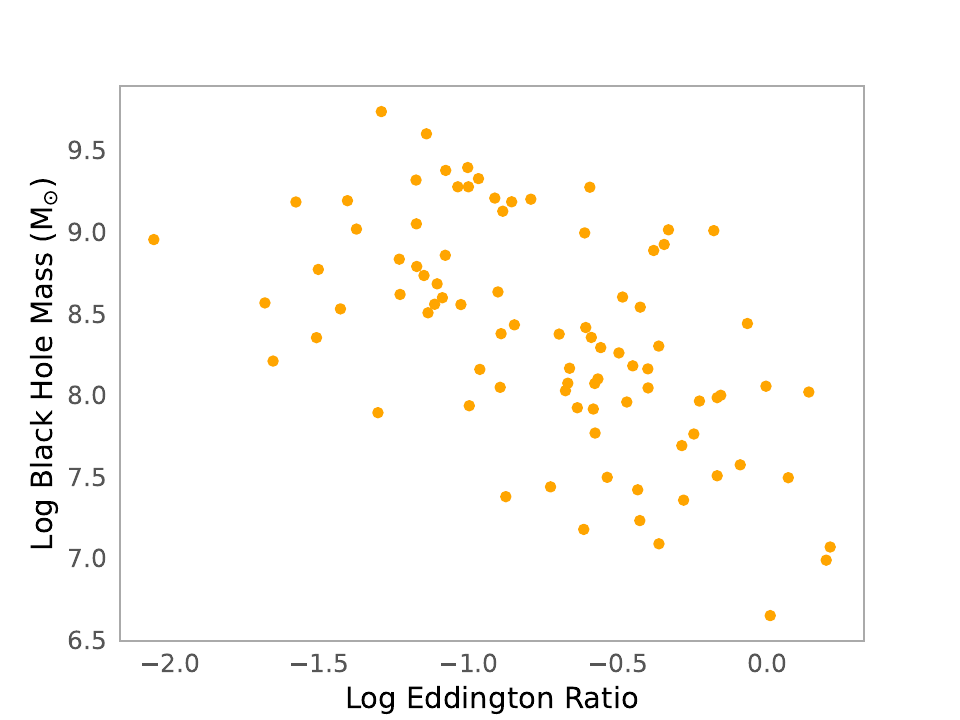}
\caption{\label{eddbhmass}Log bolometric luminosity and log black hole mass.
}

\centering
\includegraphics[scale=0.57]
{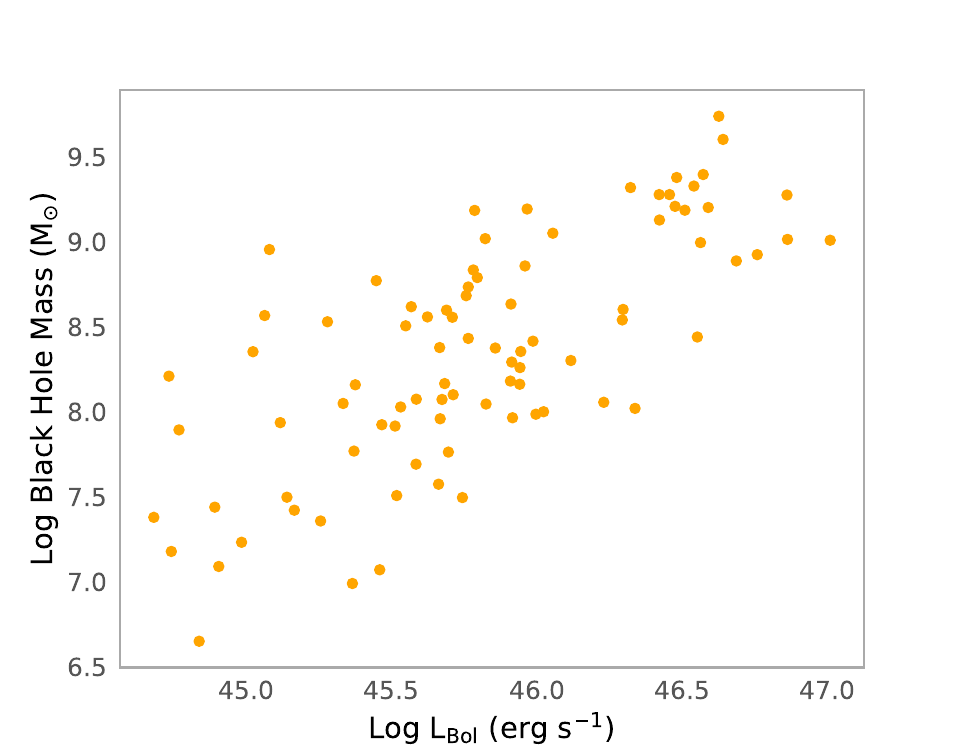}
\caption{\label{lbolbhmass}Log black hole mass and log bolometric luminosity.
}
\end{figure}

Figure \ref{figl} shows the 5100\AA\ continuum luminosity plotted against redshift $z$ for our sample as well that of Glikman et al. (2006).  Figure \ref{figboledd} plots the log bolometric luminosity against the log of the Eddington ratio for the PG quasars, while Figure \ref{eddbhmass} plots the log of the black hole mass against the log of the Eddington ratio, and Figure \ref{lbolbhmass} plots the log bolometric luminosity against the log black hole mass.

\begin{figure*}[!ht]
\centering
\includegraphics[scale=0.57]{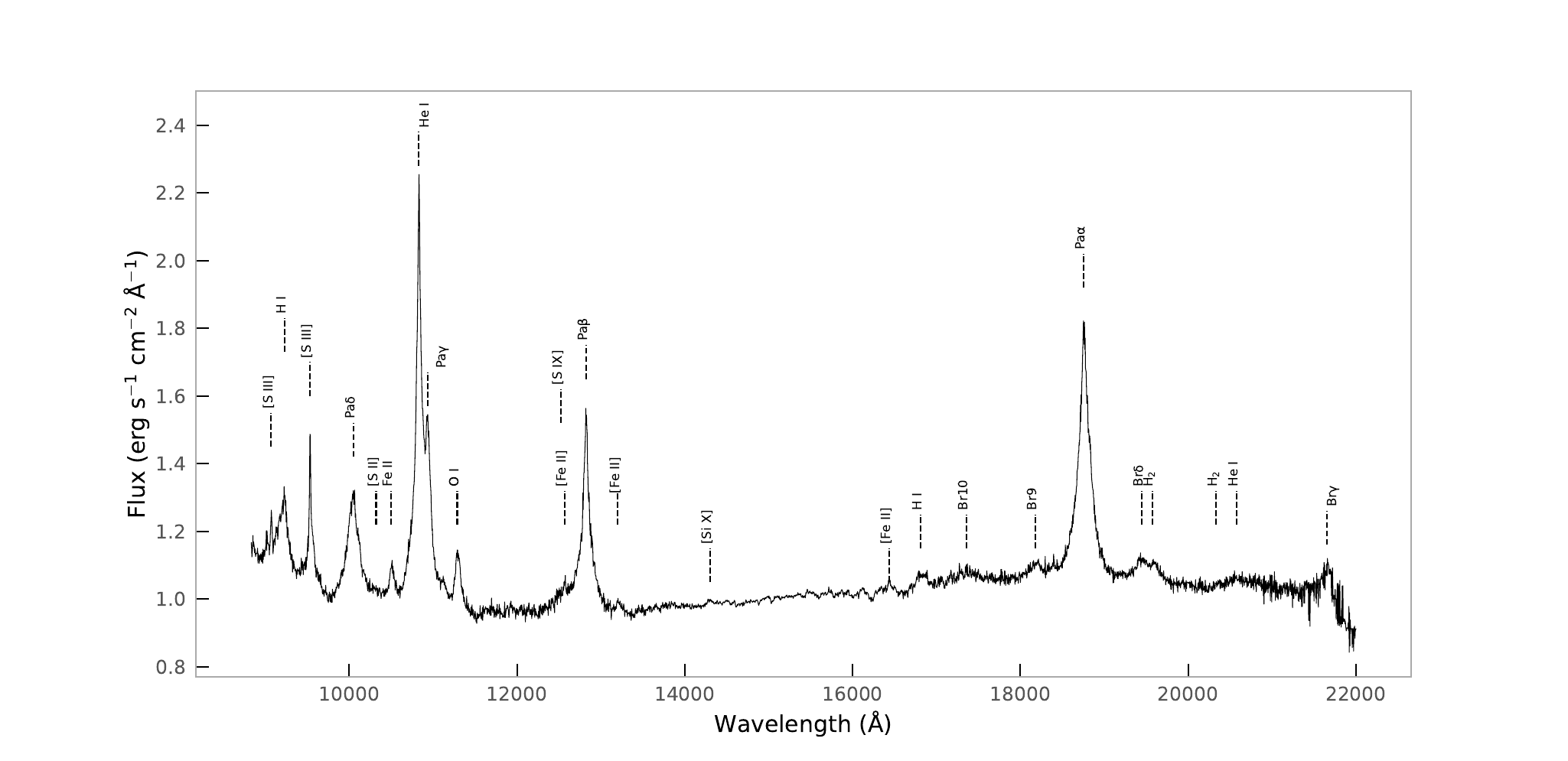}
\caption{\label{figcomp}Near-infrared median-combined composite spectrum of the $z < 0.5$ PG quasars.  The 14,000-16,0000 \AA\ range is used in the normalization process.  The flux is in normalized F$_\lambda$ units, and the wavelengths are in rest-frame Angstroms.  Notable emission lines are marked.}
\end{figure*}

\subsection{Observations}

Spectra were obtained using the Apache Point Observatory 3.5-m ARC telescope and TripleSpec near-infrared cross-dispersed spectrograph \citep{2004SPIE.5492.1295W}.  The instrument covers wavelengths of 0.95-2.46 \textmu m across five orders.  The 1.1x43" slit with a spectral resolution of R = 3500 was used.  To correct for airglow and thermal emission, spectra were taken at two positions (A and B) and differenced.  The slit position alternated in an ABBA pattern. \par
In many cases the same object was observed multiple times. 
Observations of an object across multiple nights were sometimes combined to increase the signal-to-noise ratio and this combined spectrum was used.  For other objects, the spectrum with the highest signal-to-noise ratio was used.  For telluric correction, a close standard A0V star was observed right before and sometimes after each object.  Most spectra were taken at an airmass less than 2.  See Appendix A for the observing log (Table \ref{tab:Table of Targets}). \par

\subsection{Reductions}
IDL-based Spextool (SPectral EXtraction TOOL) GUIs Xspextool \citep{2004PASP..116..362C}, Xcombspec, Xtellcor \citep{2003PASP..115..389V}, Xmergeorders, and Xcleanspec were used to reduce data.  Flat fielding, extraction, and wavelength calibration were done using the Xspextool GUI.  This GUI was used to combine the flats taken each night into a single flat.  With two of the exposures taken for each object, individual wavelength calibration and arc files were created.  Images were loaded and spectra were extracted for all standard stars and science targets.  A-B reduction mode was used in this process.  The master flat image, wavelength calibration file, and two standard star or science target images were selected.  The standard and science images were linearity-corrected, subtracted, and divided by the normalized flat image.  Average spatial profiles were taken along the slit in orders 3-7 as a median along the wavelength direction for spatial points.  After two apertures A (object) and B (negative of the object) were found in the spatial profiles, a weighted-profile method was used to extract spectra.  In this process, a 2.0 arcsecond radius was used for the extraction aperture used to find the source flux for each column.  The start radius for the background window in background subtraction was 3.0 arcseconds.  A first-order polynomial was used to fit the background and the background width was set as 2.0 arcseconds.  With these apertures defined, the spectrum was extracted.  \par  
Xspextool-extracted spectra of the same object were combined into a single spectrum using the Xcombspec GUI.  Spectra were scaled to a common flux level before being combined to account for the fact that not all spectra for each object had similar flux levels.  Order 5 was used to determine a scale factor to use on all orders of a spectrum.  Order 5 was chosen because it was a middle order.  A robust weighted mean with an 8-sigma threshold was used to combine the spectra.  In this statistic, a sigma-clipping algorithm is used to find outliers.  A weighted average of good pixels was taken. \par 
Telluric corrections were made to the combined spectra using the Xtellcor GUI.  For each science target, a standard star was used in the process.  A kernel was convolved with a model spectrum of Vega scaled to the observed standard star magnitude.  A simulated observed spectrum of the standard star with the resolution of the data resulted from this process.  To generate a telluric correction spectrum, we divided the observed standard-star spectrum by the convolved spectrum.  To correct for the effects of the Earth's atmosphere, we used the telluric correction spectrum on the object spectrum.  Due to the presence of a hydrogen line in the standard star spectrum at 1.005 \textmu m not impacted by telluric absorption, order 6 was used in this process.  A polynomial of order 11 or 12 was fit to the data to normalize the standard star spectrum. After normalization, a region that included the hydrogen line was used to create the convolution kernel.  To create the telluric spectrum, the stellar hydrogen lines were removed, and the features due to atmospheric absorption remained.  Small shifts in wavelength from the standard star spectrum to the object spectrum were found and removed for each order. \par  
The Xmergeorders GUI was used to combine individual telluric-corrected orders for each object into a continuous spectrum.  Finally, using the Xcleanspec GUI for the merged spectra, bad pixels were interpolated over, and noisy regions of the spectra were removed.  A Gaussian smoothing of the full width half maximum equal to the slit width was performed. \par
The observed H or K band spectra in three of the objects needed to be scaled due to inconsistent interband fluxes. The H band diverged from the J and K bands in the spectra of PG 1404+226 and PG 1416-129. The K band diverged from the J and H bands in the spectrum of PG 0050+124.  

We rescaled the divergent band by fitting a quadratic function and minimizing $\chi^{2}$.  For spectra in which the H band was off, a range of scale factors was defined from 0.5 to 1.5 in 0.001 increments.  Each scale factor was multiplied by the flux in the H band, and a quadratic function was fit to the H band while another quadratic function was fit to the J and K bands.  The two functions were then applied to the entire spectrum and a $\chi^{2}$ value for the two functions was obtained using the H band region.  The scale factor that minimized $\chi^{2}$ was kept and applied to the H band for the final spectrum.  For the spectrum with the K band off, the same range of scale factors was defined.  The scale factors were multiplied by the K band, and the linear functions were fit to the H and K bands and applied to all wavelengths.  The K band region was used to find $\chi^{2}$.  Again, the scale factor that minimized $\chi^{2}$ was kept and multiplied by K band for the final spectrum.  Emission lines were excluded from all fits.  Our final spectra are shown in the Appendix.\par


\section{Composite Spectra} 

To construct the composite, we started with an algorithm from GitHub \footnote{https://github.com/hklaufus/CompositeSpectrum} , but eventually made significant modifications.  We initially dereddened the spectra using the dust$\_$extinction Python package with the G23 model (although the individual spectra we plot are not dereddened).  The dereddened spectra were deredshifted by dividing the wavelengths by one plus the NED \footnote{The NASA/IPAC Extragalactic Database (NED) is funded by the National Aeronautics and Space Administration and operated by the California Institute of Technology.} preferred redshift.  They were then sorted by increasing redshift.  \par  

Normalization factors were found as follows.  For each spectrum, an initial normalization factor was found using the median flux between 14000 and 16000 \AA{} of the following spectrum divided by the median flux in the same region of the current spectrum.  This wavelength range is relatively line free and in the middle of the spectra.  This initial normalization factor was added to a normalization list.  Each element of the normalization list was multiplied by the next initial normalization factor.  The flux for each spectrum was then multiplied by the normalization factor from the normalization list.  To obtain the final composite spectrum, the normalized spectra were binned and an unweighted median was found. To do this, a new wavelength array was created that began at the minimum rest wavelength and ended at the maximum rest wavelength, with intervals of 4 \AA\ between the minimum and maximum wavelengths.  For each wavelength interval, all spectra were searched for flux values for wavelengths within that interval.  These flux values were stored in a temporary flux array, and an unweighted median was taken of the flux values in each bin.  Figure \ref{figcomp} shows the resulting composite spectrum.\

\begin{figure}[!h]
\includegraphics[scale=0.275]
{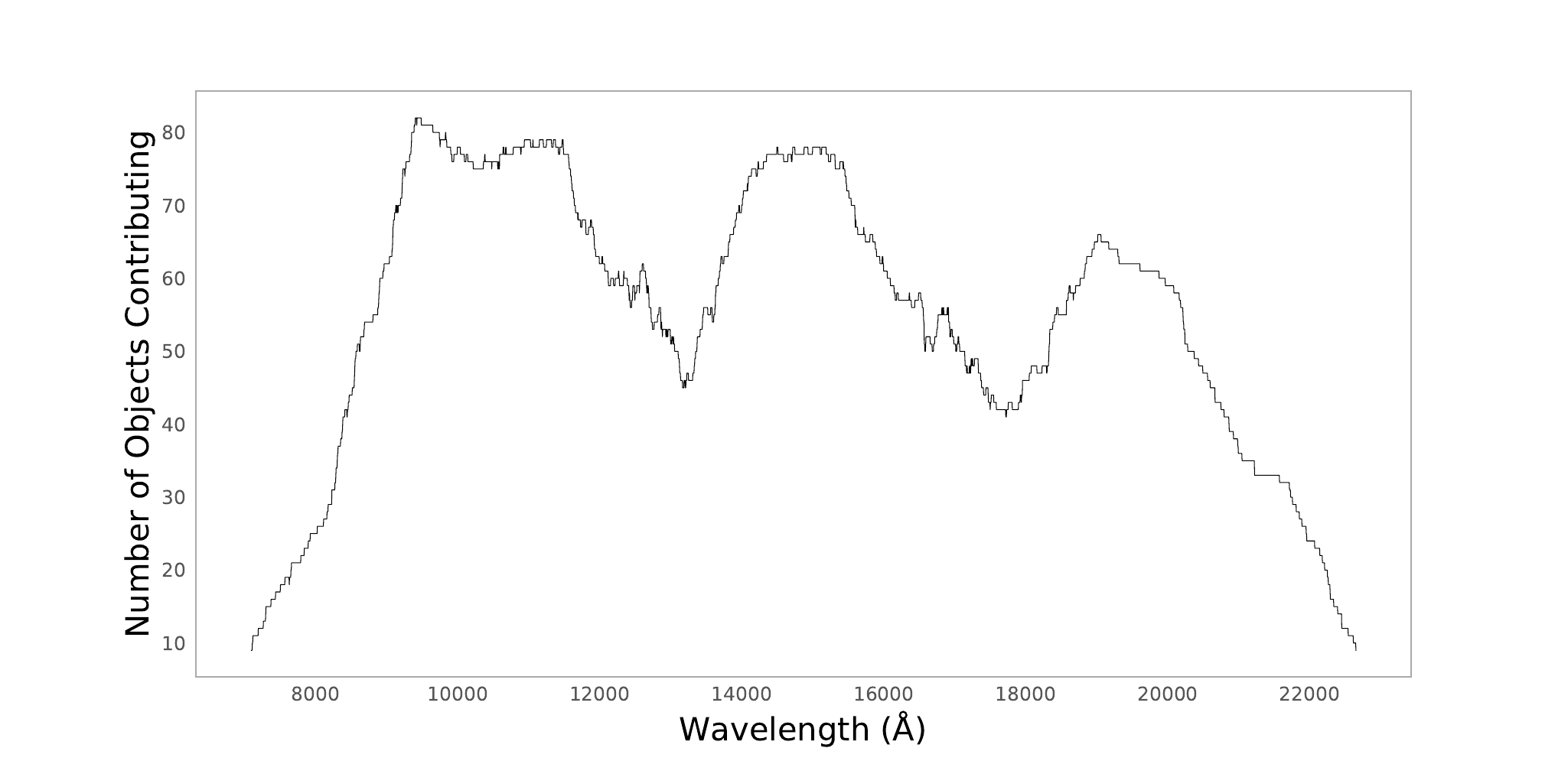}

\caption{\label{figcompnum}Number of objects contributing as a function of wavelength for the composite.  The gaps between the J, H, and K bands are due to the fact that in individual spectra, noisy regions needed to be removed between the bands.}

\includegraphics[scale=0.275]
{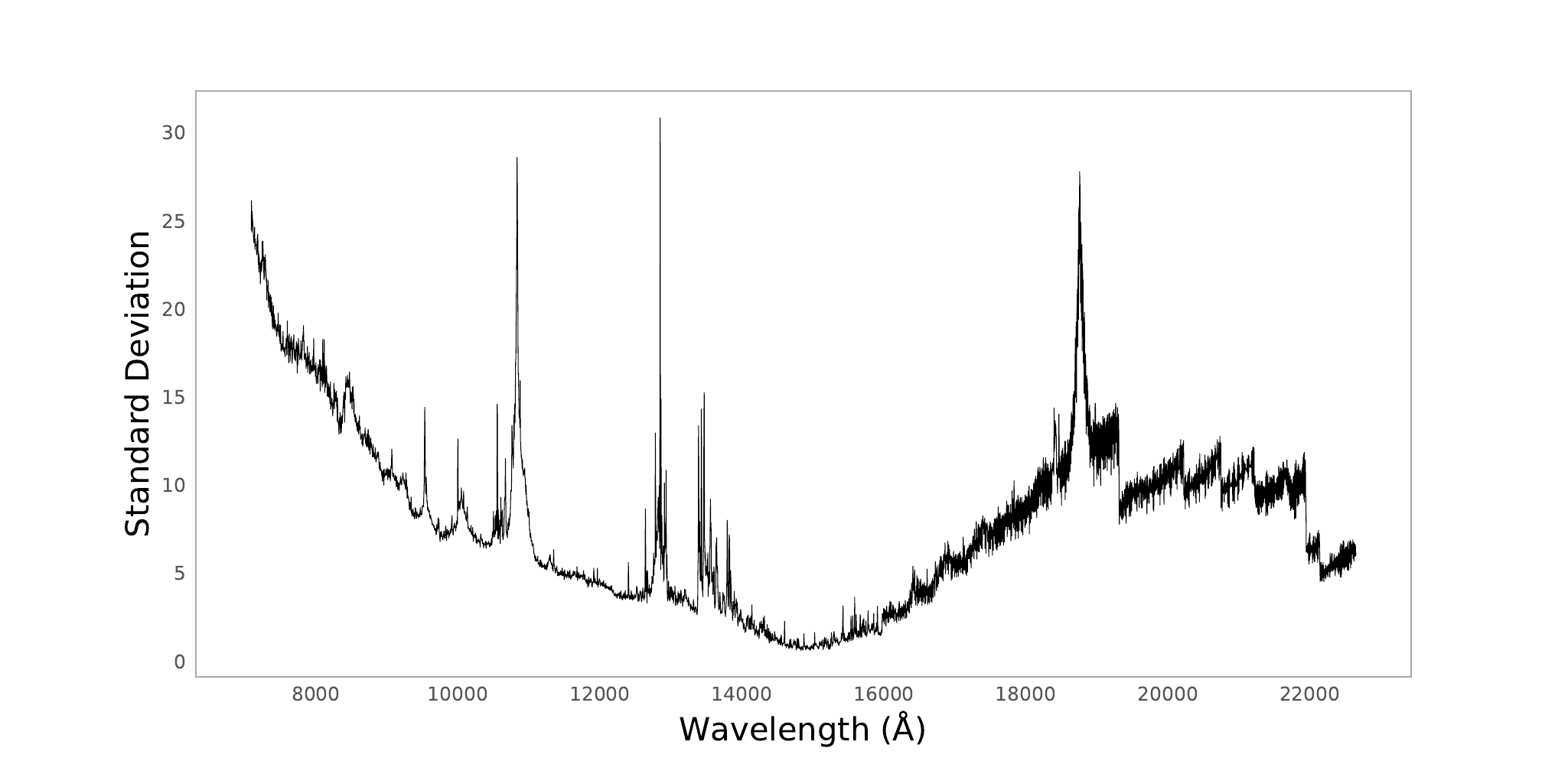}
\caption{\label{figcompstdev}Composite spectrum standard deviation.  Regions consisting of emission lines show large standard deviations, indicating variations among the contributing spectra.}

\includegraphics[scale=0.275]{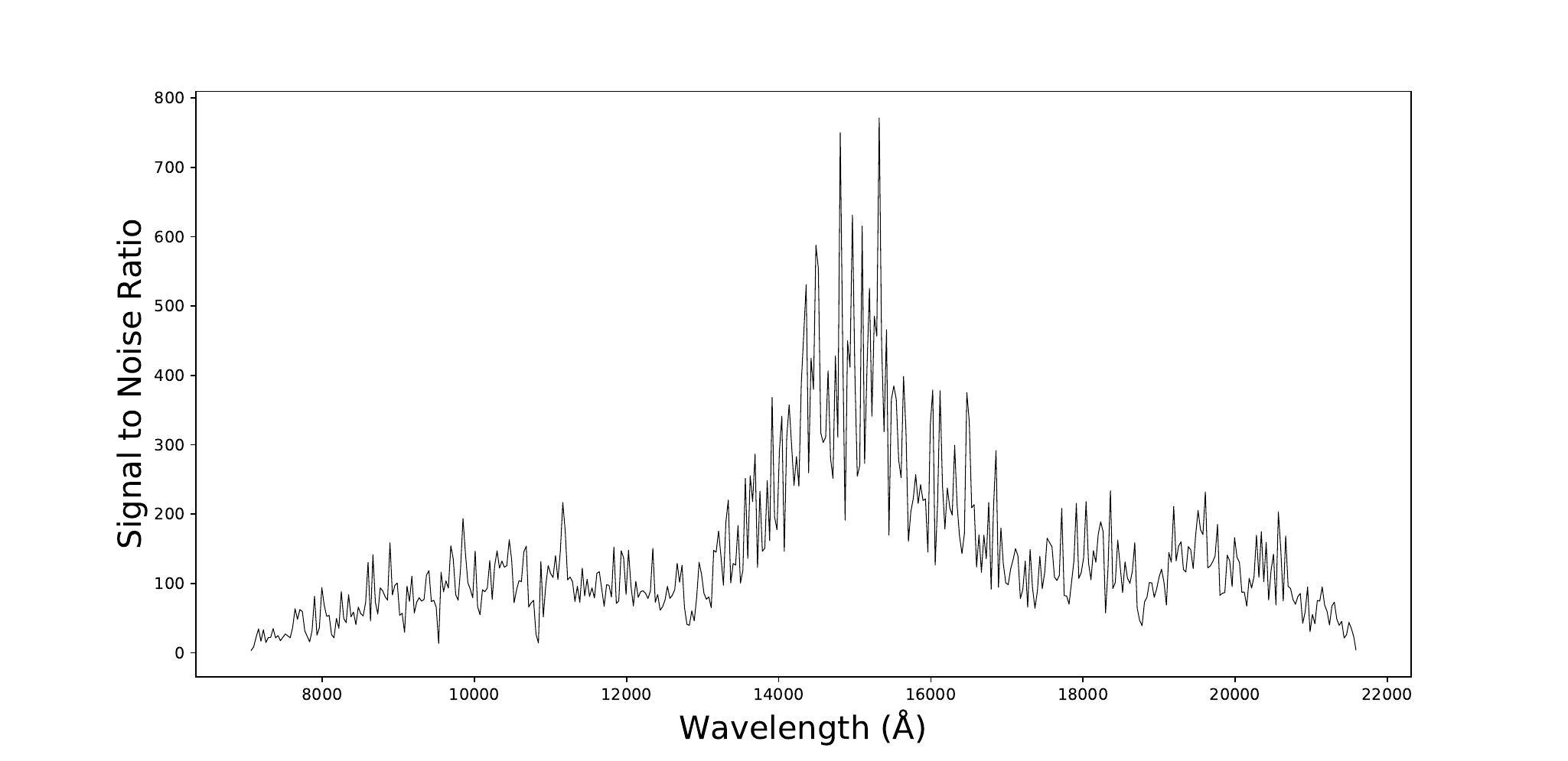}
\caption{\label{figcompsnr}Composite spectrum signal-to-noise ratio.}
\end{figure}

\begin{figure*}[!ht]
\centering
\includegraphics[scale=0.57]{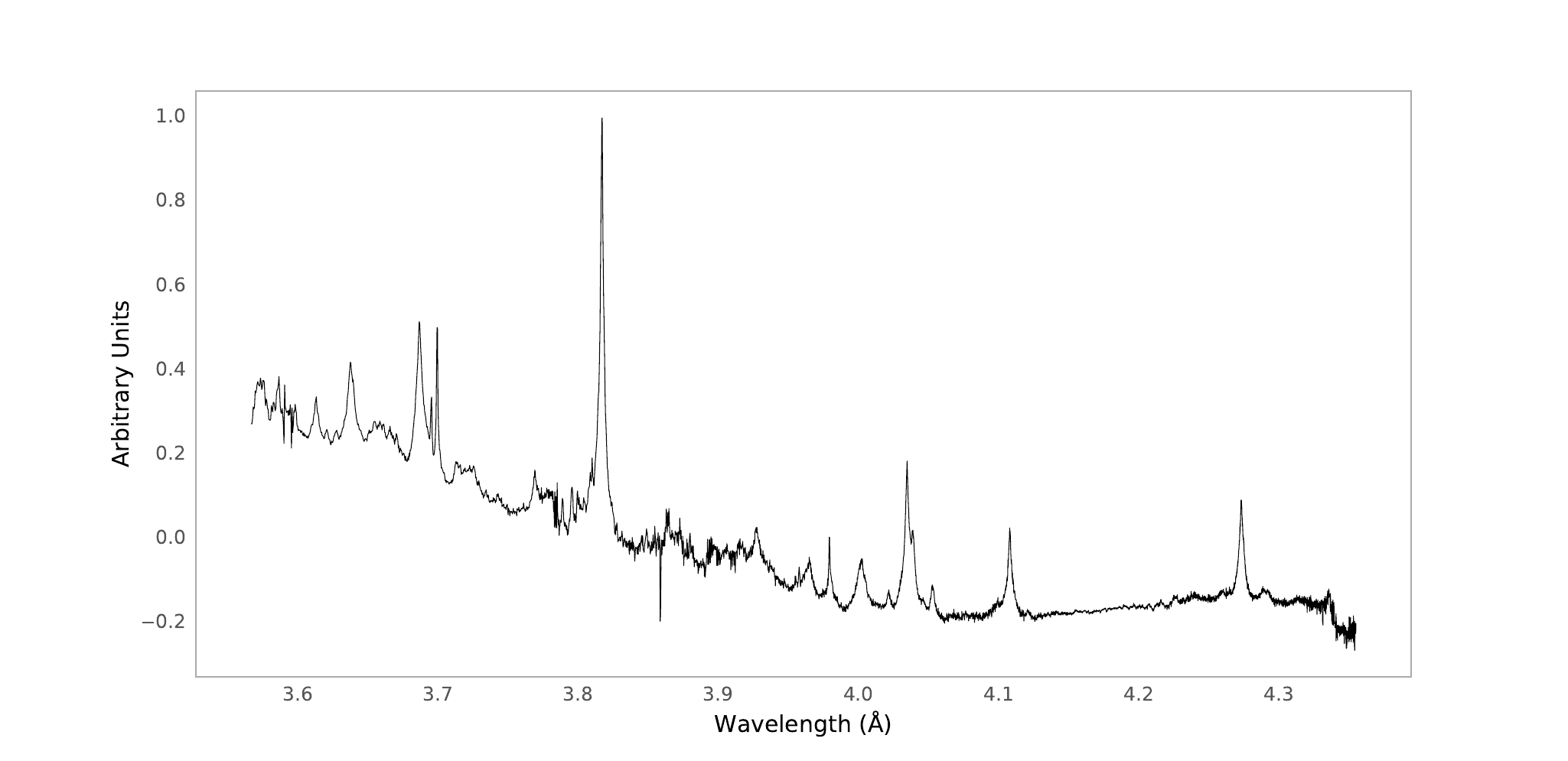}
\caption{\label{fig:optnir}Log-log combined composite spectrum using our NIR spectra and the optical spectra of \citet{1992ApJS...80..109B}.  The flux is in arbitrary F$_{\lambda}$ units, and the wavelengths are in rest-frame Angstroms}
\end{figure*}

We provide several additional figures to illustrate the characteristics of the composite spectrum.  First, Figure \ref{figcompnum} shows the number of spectra contributing to the composite as a function of wavelength.  Because our spectra have gaps between the NIR bands, and the redshift range is not large, this plot is triple-peaked.
Figure \ref{figcompstdev} shows the standard deviation spectrum; as expected given our central normalization, the standard deviation rises at both short and long wavelengths because it is where the fewest objects contribute to the composite. There are also spikes and fluctuations in the standard deviation due to emission lines and the effects of larger noise in the K band combined with contributing spectra of different objects ending at different wavelengths. 
Figure \ref{figcompsnr} shows an empirical estimate of the signal-to-noise ratio calculated over 32 \AA\ intervals.  To empirically estimate the signal-to-noise ratio, the composite spectrum was first smoothed using the Specutils function Gaussian\textunderscore
smooth with a standard deviation of 5 \citep{nicholas_earl_2023_10016569}. The spectrum was then divided by the smoothed spectrum, and the signal-to-noise ratio was estimated by dividing this normalized spectrum by the root median square noise of each 32 \AA\ interval. We note that we could have obtained an even higher signal-to-noise ratio composite spectrum by adjusting the weights of the individual spectra, but wanted to preserve the strengths of the PG sample, which is representative of heterogeneous quasar spectral properties \par.

Because it may be of interest, we also generated an optical composite spectrum of the $z < 0.5$ PG quasars using the spectra of Boroson \& Green (1992).  Because of the lack of bandpass gaps in the spectra, this is much more straightforward, although there are similar issues with limited numbers of objects at the extreme wavelengths of the optical composite.  We started with the lowest redshift spectrum and scaled subsequent spectra to the median of the overlapping spectral region, using a bin size of 3 \AA.  The resulting optical composite was then scaled to match our NIR composite spectrum.  We plot the resulting optical-NIR composite spectrum in Fig. \ref{fig:optnir}.  All of our composite spectra are available on our website (https://physics.uwyo.edu/agn/index.html).

We characterize the NIR continuum and emission lines of the composite spectrum in the appendix.

\subsection{Luminosity and Eddington Ratio Composite Spectra}

We generated many other composite spectra, and here present four of them of potential interest.  These include high and low luminosity and high and low Eddington ratio composites.  They were generated using the same process as our total PG sample NIR composite. \par 

\begin{figure*}[!ht]
\centering
\includegraphics[scale=0.57]{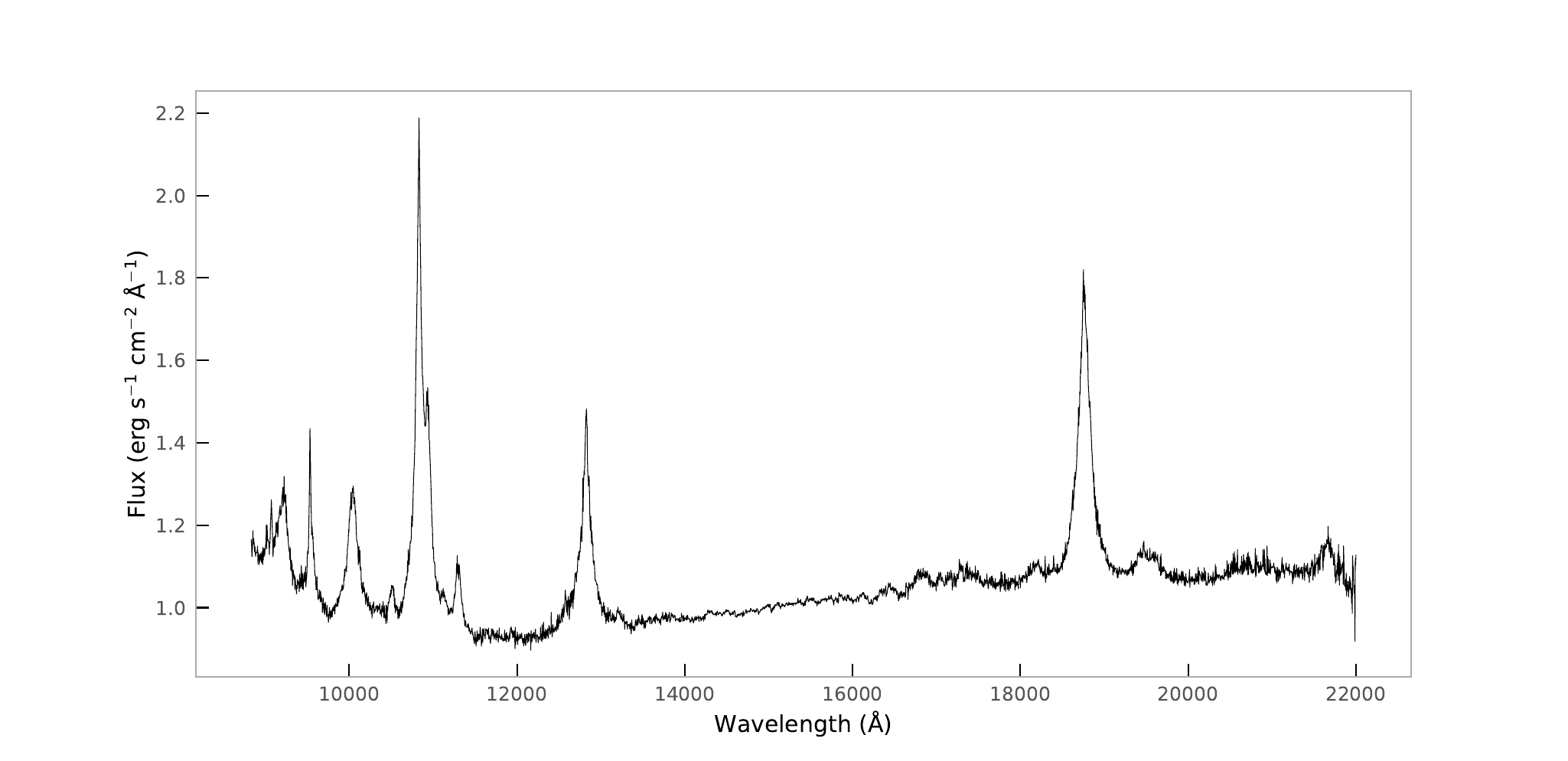}
\caption{\label{fighil}Near-infrared median-combined composite spectrum of the $z < 0.5$ PG quasars with log [$\lambda L_{\lambda}$(5100)/(erg s$^{-1}$)] $>$ 44.25.  The 14,000-16,0000 \AA\ range is used in the normalization process.  The flux is in relative F$_\lambda$ units, and the wavelengths are in rest-frame Angstroms.}

\includegraphics[scale=0.57]{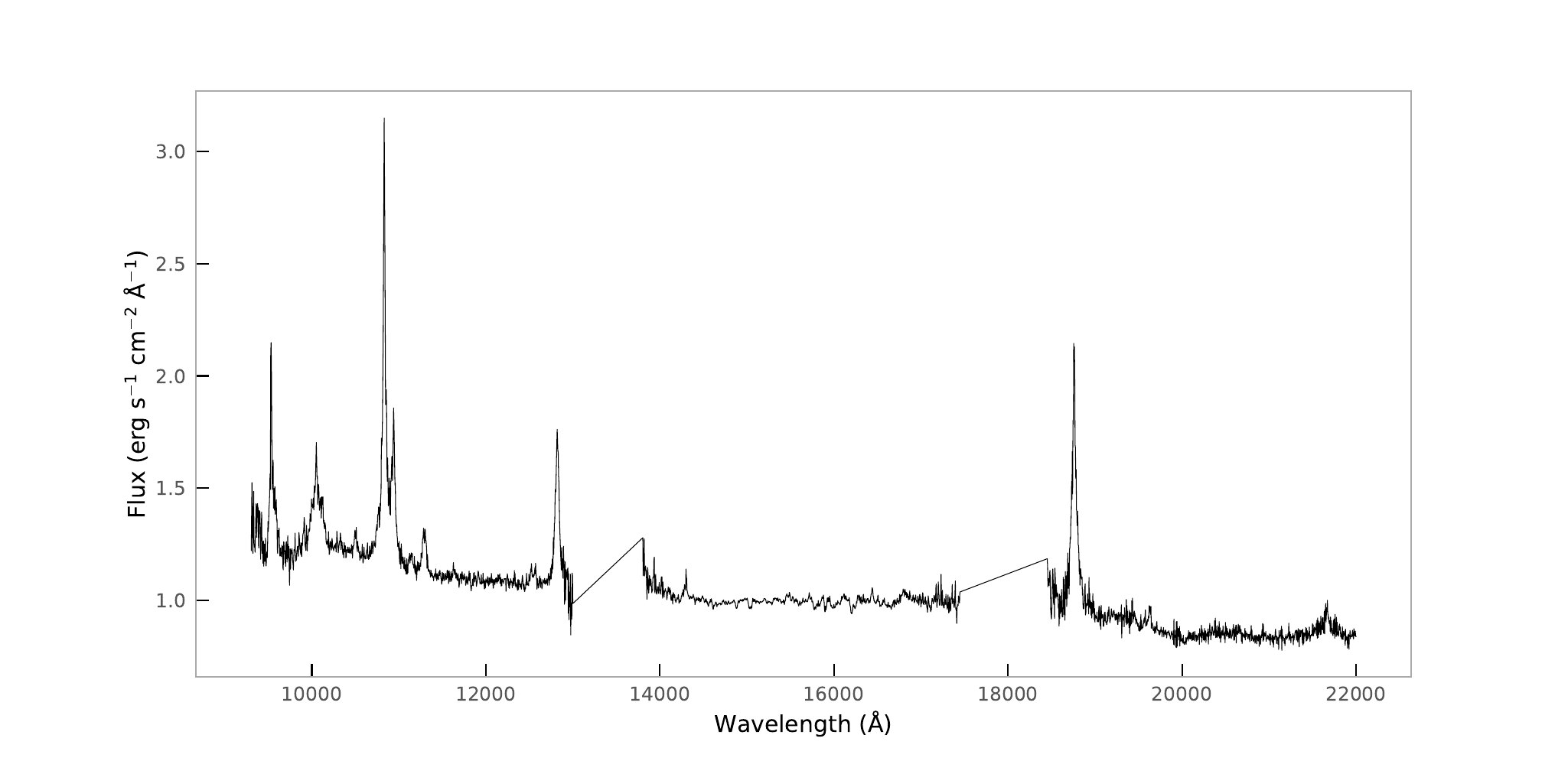}
\caption{\label{figlol}Near-infrared median-combined composite spectrum of the $z < 0.5$ PG quasars with log [$\lambda L_{\lambda}$(5100)/(erg s$^{-1}$)] $<$ 44.25.  The 14,000-16,0000 \AA\ range is used in the normalization process.  The flux is in F$_\lambda$ units, and the wavelengths are in rest-frame Angstroms.} 
\end{figure*}

First we consider luminosity.  We elected to divide the sample at log [$\lambda L_{\lambda}$(5100)/(erg s$^{-1}$)] = 44.25 as this empirically distinguishes the higher-luminosity quasar-like AGNs (Figure \ref{fighil}) from the low-luminosity and low-redshift spur of more Seyfert-like galaxies (Figure \ref{figlol}) seen in the luminosity-redshift distribution of Figure \ref{figl}.  Because of the small redshift range of the objects contributing to the low-luminosity composite, the gaps between the J, H, and K bandpasses are not filled in. 

\begin{figure*}[!ht]
\centering
\includegraphics[scale=0.57]{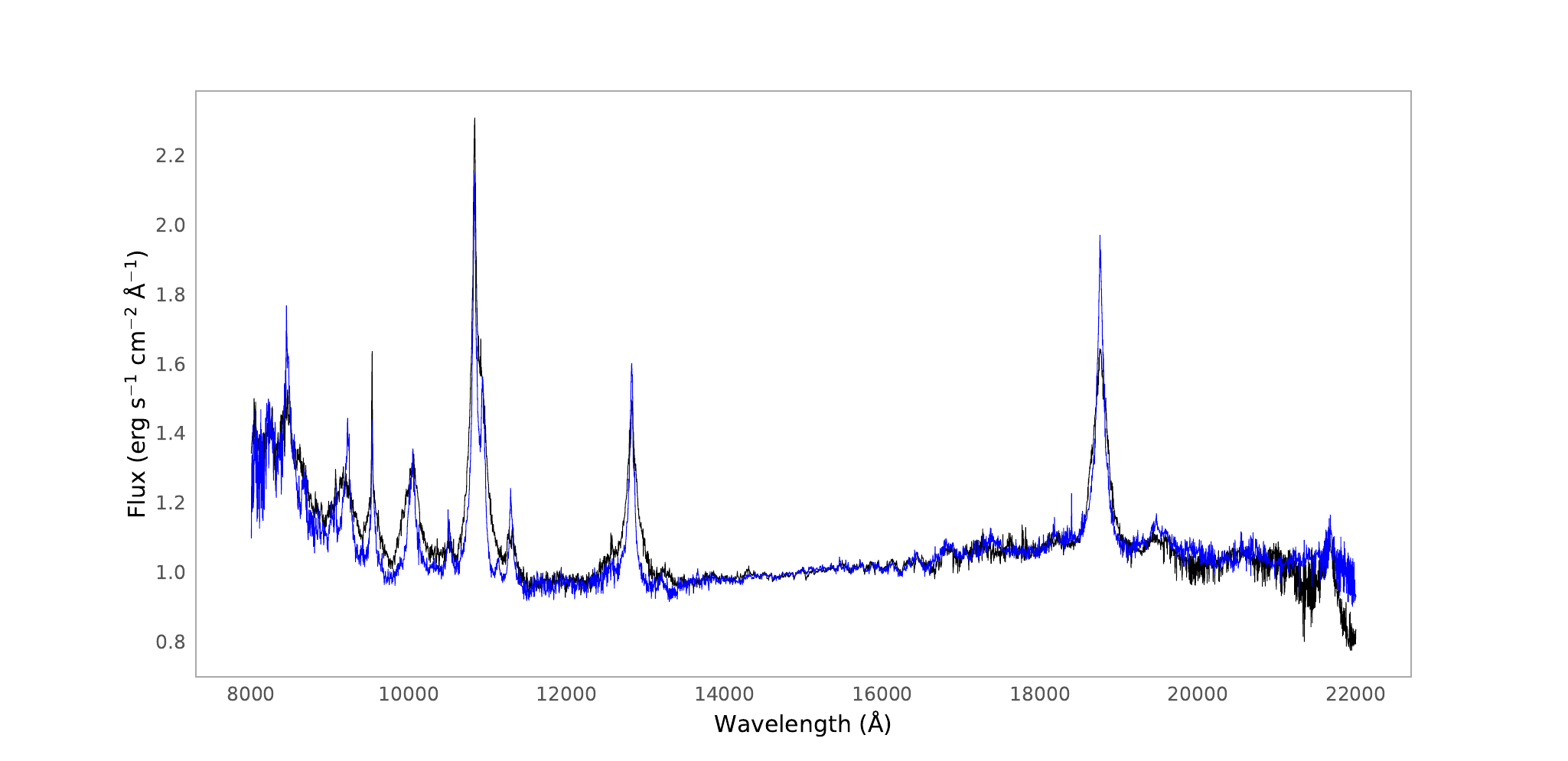}
\caption{\label{eddcomp}Composite spectra consisting of objects with Eddington ratio greater than (blue) and less than (black) the median.}

\centering
\includegraphics[scale=0.57]{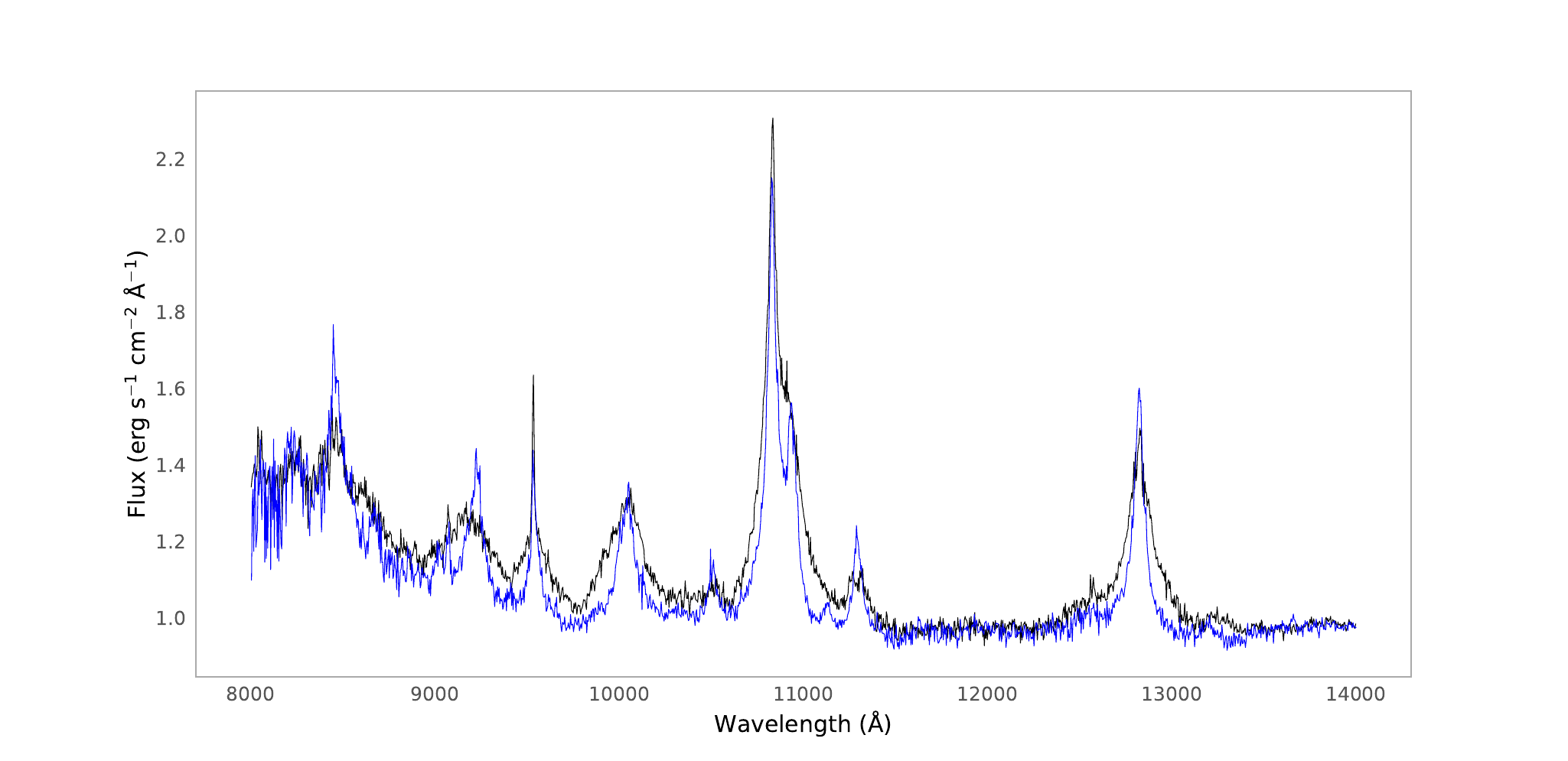}
\caption{\label{eddcompj}Composite spectra consisting of objects with Eddington ratio greater than (blue) and less than (black) the median.  The shorter wavelength range, zooming in on J band, makes the differences in the emission lines clearer than when the whole spectrum is shown.}
\end{figure*}

We take a simpler approach by presenting spectral variations with changing Eddington ratio and simply split the sample into two at the median (0.215).  Because the two composite spectra are rather similar, especially in terms of the continuum shape, we overplot them in Fig. \ref{eddcomp}.  
The clearest differences are manifested in the emission lines in the J band, which we plot by itself
in Fig. \ref{eddcompj}.
 
We experimented with the creation of separate radio-loud and radio-quiet quasar composite spectra, but there are only 17 of the former, and due to their small numbers and  disjoint distribution of their properties artificial features formed in the continuum shape.  

\section{Discussion}

\subsection{Comparison to the Glikman et al. (2006) Composite}

\begin{figure*}[!ht]
\centering
\includegraphics[scale=0.69]{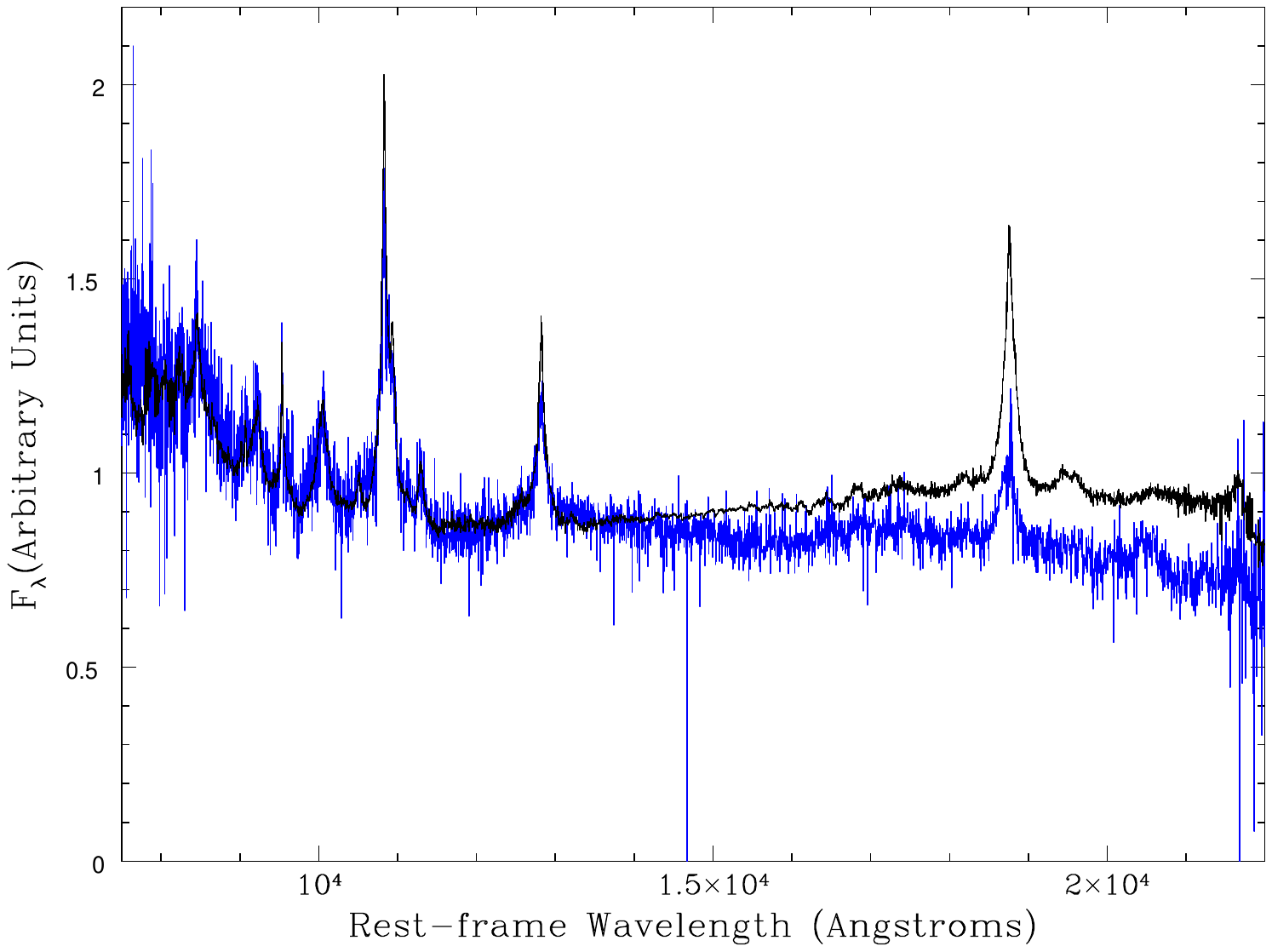}
\caption{\label{overplot}The PG quasar NIR composite spectrum (black) and the Glikman et al. (2006) NIR composite spectrum (blue) with no smoothing. Our composite has been normalized to approximately match the blue spectrum in the rest-frame J-band.}
\end{figure*}

Figure \ref{overplot} shows our PG quasar NIR composite in black and the Glikman et al. (2006) NIR composite in blue.  Our input spectra have a higher signal-to-noise ratio and we have more than three times as many objects, with similar resolution, so the quality is better and we can detect some weaker features.
The two spectra are rather similar in the emission lines to the extent
that the signal-to-noise ratio allows comparisons.  The two spectra differ significantly in terms 
of the continuum shape. Depending on the normalization, one could say that the PG composite is either deficient at short wavelengths or displays an excess at long wavelengths compared to that of the Glikman. 
Although Figure 1 shows that the PG sample includes quasars of both
lower and higher luminosity compared to the SDSS quasars used to construct the Glikman et al. (2006) composite, the optical continuum luminosity of the former is larger than that of the latter.
These continuum differences are most likely related to the differences in luminosity, which we shall discuss next.

\subsection{Luminosity Variations}

The high-luminosity composite spectrum shown in Figure \ref{fighil} closely resembles that of the total low-redshift PG sample.  The low-luminosity composite spectrum shown in Figure \ref{figlol} features several differences. Notably it has narrower broad emission lines,
likely resulting from the Seyfert-like AGN used to construct it having smaller black holes than the more luminous quasars. 
Much more striking, however, is that the low-luminosity composite displays a much bluer continuum at long wavelengths, suggestive of a much weaker and/or cooler contribution from dust emission.  

\citet{richards2006} first identified the near-to-mid infrared slope variation with their spectral energy distributions (SEDs) constructed for different luminosities \citep[see the inset for their Figure 11, as well as their follow-up paper][]{krawczyk2013}.  This hot dust bump is stronger in quasars with high luminosity.  They proposed an explanation in terms of the ``receding torus model'' 
(e.g. \citet{1991MNRAS.252..586L}), which says that higher luminosity objects have 
larger opening angles.  NIR reverberation mapping clearly supports the idea that the size of the inner torus edge, where hot dust emission is expected to dominate, correlates
tightly with luminosity 
\citep[e.g.,][]{suganuma2006,koshida2014,minezaki2021,ayubinia2025}.
The GRAVITY instrument, which has the angular resolution to see the torus directly,
confirms the correlation although finds some interesting quantitative differences
from reverberation mapping that deserve continued investigation
\citep{gravity2024}.
Richards et al. (2006) suggested 
that more edge-on views, preferentially
available in the more open high-luminosity quasars, permit better
views of the hot inner surfaces of anisotropically emitting 
clumpy torus clouds, which in turn leads to the stronger hot dust bumps in higher luminosity
quasars and the observed correlation with NIR-MIR slope.

\begin{figure*}[!ht]
\centering
\includegraphics[scale=0.57]{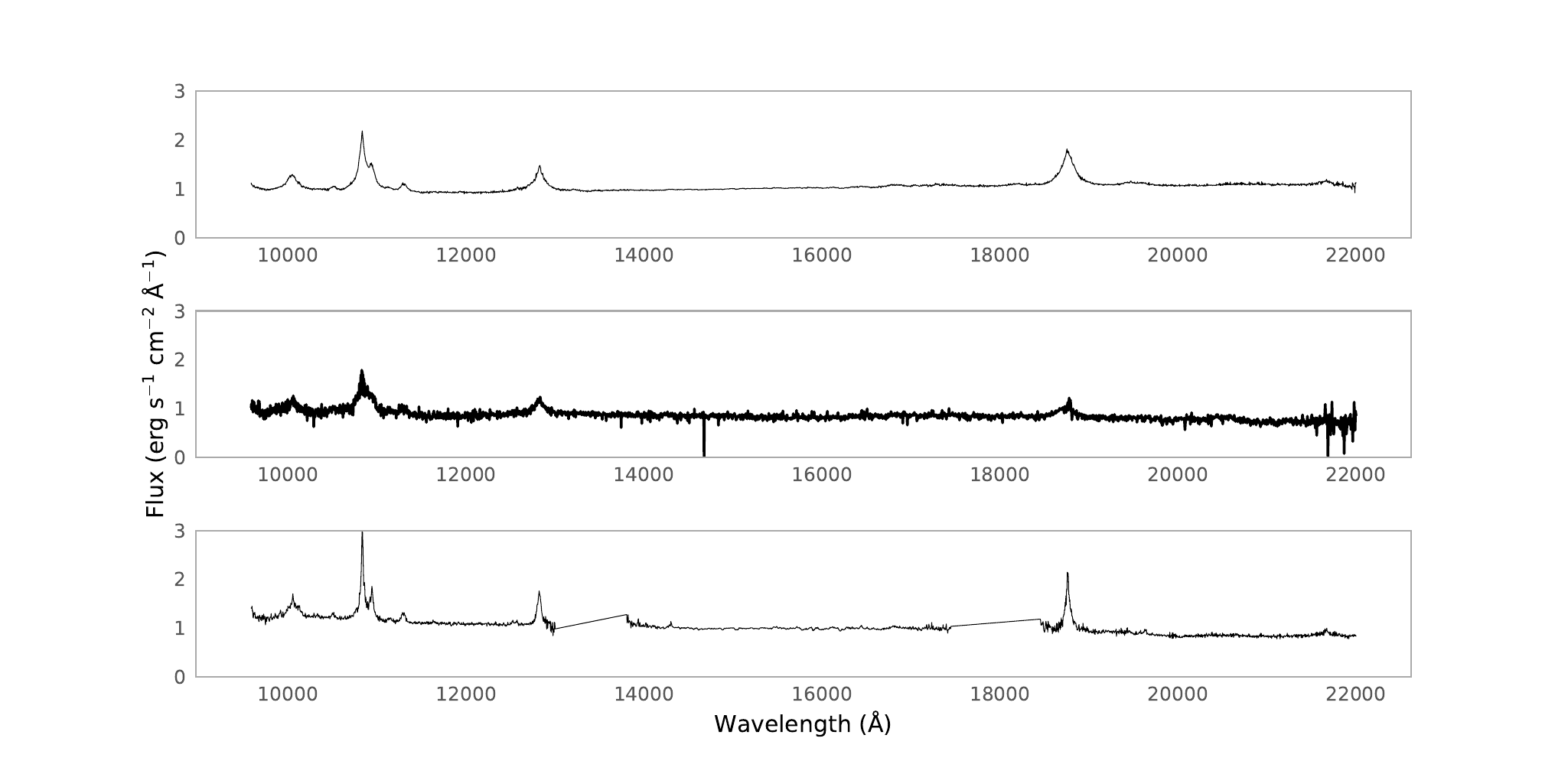}
\caption{\label{3luminosity}Composite spectra with \citet{Glikman2006} composite overplotted.  The composite consisting of spectra with log monochromatic luminosity at 5100 \AA{} greater than 44.25 is shown on top, the composite consisting of spectra with log monochromatic luminosity less than 44.25 is shown on the bottom, and the \citet{Glikman2006} composite is shown in the middle.}
\end{figure*}

Figure \ref{3luminosity} shows the Glickman et al. NIR composite sandwiched between 
our high and low-luminosity near-IR composites, all plotted on the same scale.  There 
is a trend in the long-wavelength slopes with luminosity for these composite spectra.

\subsection{Eddington Ratio Variations}
Figure \ref{eddcomp} shows very similar continuum shapes for our high and low Eddington ratio composites.  Figure \ref{eddcompj} focuses only on the J band, where some differences in the emission lines are apparent.  
The low Eddington ratio composite also shows broader broad lines in comparison to the 
high Eddington ratio composite.  This is also expected, as Figure \ref{eddbhmass} shows
that the Eddington ratio is inversely correlated with black hole mass in our PG quasars
and larger velocities are present in AGNs with larger black hole masses.

One of the clear differences at optical wavelengths is 
that low Eddington ratio objects have strong emission from the narrow-line region,
particularly [O III] \citep[e.g.,][]{sulentic2000,2002ApJ...565...78B,2014Natur.513..210S}.
There is only one strong narrow line in the NIR: [S III] $\lambda$9531.  The narrow [S III]
line is indeed stronger in the low Eddington ratio composite, as expected.
While there are Fe II emission features in the NIR that we might expect to be 
correlated with the Eddington ratio, given the EV1 trend between [O III] and optical Fe II, the NIR Fe II lines are weak and such a trend difficult to identify with composite spectra.  Still, the one relatively isolated Fe II feature blueward of the He I plus Paschen $\delta$ blend (see the Fe II template spectrum in Figure \ref{feiifit})
does appear stronger in the high Eddington composite.

According to \citet{1992ApJS...80..109B}, Eigenvector 1 is dominated by the anticorrelation between Fe II strength and [O III] strength.  Even if the fundamental physical parameter, the Eddington ratio, driving EV1 has been determined, the physical mechanisms remain unclear,
particularly for the variation in NLR emission.
Several mechanisms have been proposed. For instance,
Seyfert 1 galaxies with higher Eddington fractions have relatively softer X-ray emission 
\citep[e.g.,][]{1997ApJ...477...93L}, but not so soft as to ``turn off'' both the high and low-ionization
narrow lines  \citep[e.g.,][]{ferguson1997}.
Another idea is that dust reddening in the NLR can weaken optical lines like [O III] $\lambda$5007, but the much
longer wavelength line [O IV] 26 $\mu$m similarly weakens with increasing Eddington fraction
\citep{bian2016}. Others have suggested that the radiation pattern of ionizing photons
narrows with increasing Eddington fraction, as a ``thin'' accretion disk becomes ``slim'' 
\citep[e.g.,][]{abramowicz1988}, but this theoretical expectation remains observationally untested.
There does exist one significant result relating the Eddington fraction to a NLR property,
specifically, the density.  \citet{2012AJ....143...83X} studied a sample of Seyfert 1 galaxies from the
Sloan Digital Sky Survey and found that the electron density, based on the [S II] $\lambda$6716/$\lambda$6731
intensity ratio, decreases with increasing Eddington ratio. The effect is large, spanning
two orders of magnitude in electron density for a similar change in Eddington ratio. Given
that the forbidden lines are likely optically thin, perhaps the weakening of the NLR might
simply represent less emitting gas in the ionization cones overall. \citet{2012AJ....143...83X} suggest two
reasons why there might be lower density gas in the most highly accreting Seyfert galaxies:
bar-driven inflows replenishing the NLR with low-density gas, or radiation-pressure-driven
disk winds, which would not be unexpected.

\section{Summary}

This work presents a high signal-to-noise ratio near-infrared median composite spectrum constructed using the 87 PG with $ z <0.5$\ quasars with optical spectra previously analyzed by \citet{1992ApJS...80..109B}.  Our spectrum complements the NIR composite of \citet{Glikman2006} for wavelengths below 2.2 $\mu$ m.  We fit a combined power law and blackbody to our NIR composite continuum, and report measurements of the stronger emission lines.  We also constructed composites of high and low luminosity subsamples as well as high and low Eddington ratio subsamples.  The clearest difference with luminosity is that the
hot dust emission clearly increases with luminosity, consistent with previous SED studies, as does the velocity width of the broad lines due to increases in black hole mass.
While spectral variations are strong in the optical as Eddingtion ratio changes, 
particularly involving the inverse correlation between the narrow [O III] line and the 
broad Fe II blends, the relative paucity of narrow lines and Fe II lines in the NIR 
means rather similar spectra in the high and low Eddington ratio composites.

\par

\begin{acknowledgments}

This research has made use of the NASA/IPAC Extragalactic Database (NED), which is funded by the National Aeronautics and Space Administration and operated by the California Institute of Technology. \par
AC was funded by Wyoming NASA Space Grant Consortium, NASA Grant NNX15AI08H. \par
We also thank Patrick and Nora Ivers for their generous donation partially supporting AC, as well as the Apache Point Observatory telescope operators for their help with the observations.
Based on observations obtained with the Apache Point Observatory 3.5-meter telescope, which is owned and operated by the Astrophysical Research Consortium.

\end{acknowledgments}

\bibliographystyle{aasjournal}
\bibliography{citations.bib}

\begin{thebibliography}{}
\expandafter\ifx\csname natexlab\endcsname\relax\def\natexlab#1{#1}\fi
\providecommand{\url}[1]{\href{#1}{#1}}
\providecommand{\dodoi}[1]{doi:~\href{http://doi.org/#1}{\nolinkurl{#1}}}
\providecommand{\doeprint}[1]{\href{http://ascl.net/#1}{\nolinkurl{http://ascl.net/#1}}}
\providecommand{\doarXiv}[1]{\href{https://arxiv.org/abs/#1}{\nolinkurl{https://arxiv.org/abs/#1}}}

\bibitem[{{Abramowicz} {et~al.}(1988){Abramowicz}, {Czerny}, {Lasota}, \& {Szuszkiewicz}}]{abramowicz1988}
{Abramowicz}, M.~A., {Czerny}, B., {Lasota}, J.~P., \& {Szuszkiewicz}, E. 1988, \apj, 332, 646, \dodoi{10.1086/166683}

\bibitem[{{Ayubinia} {et~al.}(2025){Ayubinia}, {Woo}, {Wang}, {Mandal}, \& {Son}}]{ayubinia2025}
{Ayubinia}, A., {Woo}, J.-H., {Wang}, S., {Mandal}, A.~K., \& {Son}, D. 2025, \apj, 994, 188, \dodoi{10.3847/1538-4357/ae101f}

\bibitem[{{Bian} {et~al.}(2016){Bian}, {He}, {Green}, {Shi}, {Ge}, \& {Liu}}]{bian2016}
{Bian}, W.-H., {He}, Z.-C., {Green}, R., {et~al.} 2016, \mnras, 456, 4081, \dodoi{10.1093/mnras/stv2936}

\bibitem[{{Boroson}(2002)}]{2002ApJ...565...78B}
{Boroson}, T.~A. 2002, \apj, 565, 78, \dodoi{10.1086/324486}

\bibitem[{{Boroson} \& {Green}(1992)}]{1992ApJS...80..109B}
{Boroson}, T.~A., \& {Green}, R.~F. 1992, \apjs, 80, 109, \dodoi{10.1086/191661}

\bibitem[{{Cushing} {et~al.}(2004){Cushing}, {Vacca}, \& {Rayner}}]{2004PASP..116..362C}
{Cushing}, M.~C., {Vacca}, W.~D., \& {Rayner}, J.~T. 2004, \pasp, 116, 362, \dodoi{10.1086/382907}

\bibitem[{Earl {et~al.}(2023)Earl, Tollerud, O'Steen, brechmos, Kerzendorf, Busko, shaileshahuja, D'Avella, Robitaille, Lim, Ginsburg, Homeier, Sipőcz, Averbukh, Tocknell, Cherinka, Ogaz, Geda, Davies, Conroy, Günther, Barbary, Foster, Droettboom, Nguyen, Bray, Casey, Teuben, Crawford, \& Ferguson}]{nicholas_earl_2023_10016569}
Earl, N., Tollerud, E., O'Steen, R., {et~al.} 2023, \dodoi{10.5281/zenodo.10016569}

\bibitem[{{Ferguson} {et~al.}(1997){Ferguson}, {Korista}, {Baldwin}, \& {Ferland}}]{ferguson1997}
{Ferguson}, J.~W., {Korista}, K.~T., {Baldwin}, J.~A., \& {Ferland}, G.~J. 1997, \apj, 487, 122, \dodoi{10.1086/304611}

\bibitem[{{Garcia-Rissmann} {et~al.}(2012){Garcia-Rissmann}, {Rodr{\'\i}guez-Ardila}, {Sigut}, \& {Pradhan}}]{2012ApJ...751....7G}
{Garcia-Rissmann}, A., {Rodr{\'\i}guez-Ardila}, A., {Sigut}, T.~A.~A., \& {Pradhan}, A.~K. 2012, \apj, 751, 7, \dodoi{10.1088/0004-637X/751/1/7}

\bibitem[{{Glikman} {et~al.}(2006){Glikman}, {Helfand}, \& {White}}]{Glikman2006}
{Glikman}, E., {Helfand}, D.~J., \& {White}, R.~L. 2006, \apj, 640, 579, \dodoi{10.1086/500098}

\bibitem[{{GRAVITY Collaboration} {et~al.}(2024){GRAVITY Collaboration}, {Amorim}, {Bourdarot}, {Brandner}, {Cao}, {Cl{\'e}net}, {Davies}, {de Zeeuw}, {Dexter}, {Drescher}, {Eckart}, {Eisenhauer}, {Fabricius}, {Feuchtgruber}, {F{\"o}rster Schreiber}, {Garcia}, {Genzel}, {Gillessen}, {Gratadour}, {H{\"o}nig}, {Kishimoto}, {Lacour}, {Lutz}, {Millour}, {Netzer}, {Ott}, {Perraut}, {Perrin}, {Peterson}, {Petrucci}, {Pfuhl}, {Prieto}, {Rabien}, {Rouan}, {Santos}, {Shangguan}, {Shimizu}, {Sternberg}, {Straubmeier}, {Sturm}, {Tacconi}, {Tristram}, {Widmann}, \& {Woillez}}]{gravity2024}
{GRAVITY Collaboration}, {Amorim}, A., {Bourdarot}, G., {et~al.} 2024, \aap, 690, A76, \dodoi{10.1051/0004-6361/202450746}

\bibitem[{{Jester} {et~al.}(2005){Jester}, {Schneider}, {Richards}, {Green}, {Schmidt}, {Hall}, {Strauss}, {Vanden Berk}, {Stoughton}, {Gunn}, {Brinkmann}, {Kent}, {Smith}, {Tucker}, \& {Yanny}}]{jester2005}
{Jester}, S., {Schneider}, D.~P., {Richards}, G.~T., {et~al.} 2005, \aj, 130, 873, \dodoi{10.1086/432466}

\bibitem[{{Kellermann} {et~al.}(1989){Kellermann}, {Sramek}, {Schmidt}, {Shaffer}, \& {Green}}]{1989AJ.....98.1195K}
{Kellermann}, K.~I., {Sramek}, R., {Schmidt}, M., {Shaffer}, D.~B., \& {Green}, R. 1989, \aj, 98, 1195, \dodoi{10.1086/115207}

\bibitem[{{Kim} {et~al.}(2018){Kim}, {Im}, {Canalizo}, {Kim}, {Kim}, {Woo}, {Taak}, {Kim}, \& {Lazarova}}]{2018ApJS..238...37K}
{Kim}, D., {Im}, M., {Canalizo}, G., {et~al.} 2018, \apjs, 238, 37, \dodoi{10.3847/1538-4365/aadfd5}

\bibitem[{{Kishimoto} {et~al.}(2008){Kishimoto}, {Antonucci}, {Blaes}, {Lawrence}, {Boisson}, {Albrecht}, \& {Leipski}}]{2008Natur.454..492K}
{Kishimoto}, M., {Antonucci}, R., {Blaes}, O., {et~al.} 2008, \nat, 454, 492, \dodoi{10.1038/nature07114}

\bibitem[{{Koshida} {et~al.}(2014){Koshida}, {Minezaki}, {Yoshii}, {Kobayashi}, {Sakata}, {Sugawara}, {Enya}, {Suganuma}, {Tomita}, {Aoki}, \& {Peterson}}]{koshida2014}
{Koshida}, S., {Minezaki}, T., {Yoshii}, Y., {et~al.} 2014, \apj, 788, 159, \dodoi{10.1088/0004-637X/788/2/159}

\bibitem[{{Krawczyk} {et~al.}(2013){Krawczyk}, {Richards}, {Mehta}, {Vogeley}, {Gallagher}, {Leighly}, {Ross}, \& {Schneider}}]{krawczyk2013}
{Krawczyk}, C.~M., {Richards}, G.~T., {Mehta}, S.~S., {et~al.} 2013, \apjs, 206, 4, \dodoi{10.1088/0067-0049/206/1/4}

\bibitem[{{Krolik}(1999)}]{1999agnc.book.....K}
{Krolik}, J.~H. 1999, {Active galactic nuclei : from the central black hole to the galactic environment}

\bibitem[{{Landt} {et~al.}(2011){Landt}, {Elvis}, {Ward}, {Bentz}, {Korista}, \& {Karovska}}]{2011MNRAS.414..218L}
{Landt}, H., {Elvis}, M., {Ward}, M.~J., {et~al.} 2011, \mnras, 414, 218, \dodoi{10.1111/j.1365-2966.2011.18383.x}

\bibitem[{{Laor} {et~al.}(1997){Laor}, {Fiore}, {Elvis}, {Wilkes}, \& {McDowell}}]{1997ApJ...477...93L}
{Laor}, A., {Fiore}, F., {Elvis}, M., {Wilkes}, B.~J., \& {McDowell}, J.~C. 1997, \apj, 477, 93, \dodoi{10.1086/303696}

\bibitem[{{Lawrence}(1991)}]{1991MNRAS.252..586L}
{Lawrence}, A. 1991, \mnras, 252, 586, \dodoi{10.1093/mnras/252.4.586}

\bibitem[{{Minezaki} {et~al.}(2019){Minezaki}, {Yoshii}, {Kobayashi}, {Sugawara}, {Sakata}, {Enya}, {Koshida}, {Tomita}, {Suganuma}, {Aoki}, \& {Peterson}}]{minezaki2021}
{Minezaki}, T., {Yoshii}, Y., {Kobayashi}, Y., {et~al.} 2019, \apj, 886, 150, \dodoi{10.3847/1538-4357/ab4f7b}

\bibitem[{{Netzer} {et~al.}(2007){Netzer}, {Lutz}, {Schweitzer}, {Contursi}, {Sturm}, {Tacconi}, {Veilleux}, {Kim}, {Rupke}, {Baker}, {Dasyra}, {Mazzarella}, \& {Lord}}]{2007ApJ...666..806N}
{Netzer}, H., {Lutz}, D., {Schweitzer}, M., {et~al.} 2007, \apj, 666, 806, \dodoi{10.1086/520716}

\bibitem[{{Petric} {et~al.}(2015){Petric}, {Ho}, {Flagey}, \& {Scoville}}]{2015ApJS..219...22P}
{Petric}, A.~O., {Ho}, L.~C., {Flagey}, N. J.~M., \& {Scoville}, N.~Z. 2015, \apjs, 219, 22, \dodoi{10.1088/0067-0049/219/2/22}

\bibitem[{{Piconcelli} {et~al.}(2005){Piconcelli}, {Jimenez-Bail{\'o}n}, {Guainazzi}, {Schartel}, {Rodr{\'\i}guez-Pascual}, \& {Santos-Lle{\'o}}}]{2005A&A...432...15P}
{Piconcelli}, E., {Jimenez-Bail{\'o}n}, E., {Guainazzi}, M., {et~al.} 2005, \aap, 432, 15, \dodoi{10.1051/0004-6361:20041621}

\bibitem[{{Richards} {et~al.}(2006){Richards}, {Lacy}, {Storrie-Lombardi}, {Hall}, {Gallagher}, {Hines}, {Fan}, {Papovich}, {Vanden Berk}, {Trammell}, {Schneider}, {Vestergaard}, {York}, {Jester}, {Anderson}, {Budav{\'a}ri}, \& {Szalay}}]{richards2006}
{Richards}, G.~T., {Lacy}, M., {Storrie-Lombardi}, L.~J., {et~al.} 2006, \apjs, 166, 470, \dodoi{10.1086/506525}

\bibitem[{{Riffel} {et~al.}(2006){Riffel}, {Rodr{\'\i}guez-Ardila}, \& {Pastoriza}}]{2006A&A...457...61R}
{Riffel}, R., {Rodr{\'\i}guez-Ardila}, A., \& {Pastoriza}, M.~G. 2006, \aap, 457, 61, \dodoi{10.1051/0004-6361:20065291}

\bibitem[{{Runnoe} {et~al.}(2012){Runnoe}, {Brotherton}, \& {Shang}}]{2012MNRAS.426.2677R}
{Runnoe}, J.~C., {Brotherton}, M.~S., \& {Shang}, Z. 2012, \mnras, 426, 2677, \dodoi{10.1111/j.1365-2966.2012.21644.x}

\bibitem[{{Sanders} {et~al.}(2020){Sanders}, {Jones}, {Shapley}, {Reddy}, {Kriek}, {Coil}, {Siana}, {Mobasher}, {Shivaei}, {Price}, {Freeman}, {Azadi}, {Leung}, {Fetherolf}, {Zick}, {de Groot}, {Barro}, \& {Fornasini}}]{2020ApJ...888L..11S}
{Sanders}, R.~L., {Jones}, T., {Shapley}, A.~E., {et~al.} 2020, \apjl, 888, L11, \dodoi{10.3847/2041-8213/ab5d40}

\bibitem[{{Schmidt} \& {Green}(1983)}]{1983ApJ...269..352S}
{Schmidt}, M., \& {Green}, R.~F. 1983, \apj, 269, 352, \dodoi{10.1086/161048}

\bibitem[{{Shang} {et~al.}(2007){Shang}, {Wills}, {Wills}, \& {Brotherton}}]{2007AJ....134..294S}
{Shang}, Z., {Wills}, B.~J., {Wills}, D., \& {Brotherton}, M.~S. 2007, \aj, 134, 294, \dodoi{10.1086/518505}

\bibitem[{{Shen} \& {Ho}(2014)}]{2014Natur.513..210S}
{Shen}, Y., \& {Ho}, L.~C. 2014, \nat, 513, 210, \dodoi{10.1038/nature13712}

\bibitem[{{Shi} {et~al.}(2014){Shi}, {Rieke}, {Ogle}, {Su}, \& {Balog}}]{2014ApJS..214...23S}
{Shi}, Y., {Rieke}, G.~H., {Ogle}, P.~M., {Su}, K.~Y.~L., \& {Balog}, Z. 2014, \apjs, 214, 23, \dodoi{10.1088/0067-0049/214/2/23}

\bibitem[{{Suganuma} {et~al.}(2006){Suganuma}, {Yoshii}, {Kobayashi}, {Minezaki}, {Enya}, {Tomita}, {Aoki}, {Koshida}, \& {Peterson}}]{suganuma2006}
{Suganuma}, M., {Yoshii}, Y., {Kobayashi}, Y., {et~al.} 2006, \apj, 639, 46, \dodoi{10.1086/499326}

\bibitem[{{Sulentic} {et~al.}(2000){Sulentic}, {Zwitter}, {Marziani}, \& {Dultzin-Hacyan}}]{sulentic2000}
{Sulentic}, J.~W., {Zwitter}, T., {Marziani}, P., \& {Dultzin-Hacyan}, D. 2000, \apjl, 536, L5, \dodoi{10.1086/312717}

\bibitem[{{Sun} \& {Shen}(2015)}]{2015ApJ...804L..15S}
{Sun}, J., \& {Shen}, Y. 2015, \apjl, 804, L15, \dodoi{10.1088/2041-8205/804/1/L15}

\bibitem[{{Vacca} {et~al.}(2003){Vacca}, {Cushing}, \& {Rayner}}]{2003PASP..115..389V}
{Vacca}, W.~D., {Cushing}, M.~C., \& {Rayner}, J.~T. 2003, \pasp, 115, 389, \dodoi{10.1086/346193}

\bibitem[{{Vestergaard} \& {Peterson}(2006)}]{2006ApJ...641..689V}
{Vestergaard}, M., \& {Peterson}, B.~M. 2006, \apj, 641, 689, \dodoi{10.1086/500572}

\bibitem[{{Wills} {et~al.}(1999){Wills}, {Laor}, {Brotherton}, {Wills}, {Wilkes}, {Ferland}, \& {Shang}}]{1999ApJ...515L..53W}
{Wills}, B.~J., {Laor}, A., {Brotherton}, M.~S., {et~al.} 1999, \apjl, 515, L53, \dodoi{10.1086/311980}

\bibitem[{{Wilson} {et~al.}(2004){Wilson}, {Henderson}, {Herter}, {Matthews}, {Skrutskie}, {Adams}, {Moon}, {Smith}, {Gautier}, {Ressler}, {Soifer}, {Lin}, {Howard}, {LaMarr}, {Stolberg}, \& {Zink}}]{2004SPIE.5492.1295W}
{Wilson}, J.~C., {Henderson}, C.~P., {Herter}, T.~L., {et~al.} 2004, in Society of Photo-Optical Instrumentation Engineers (SPIE) Conference Series, Vol. 5492, Ground-based Instrumentation for Astronomy, ed. A.~F.~M. {Moorwood} \& M.~{Iye}, 1295--1305, \dodoi{10.1117/12.550925}

\bibitem[{{Wisotzki} {et~al.}(2000){Wisotzki}, {Christlieb}, {Bade}, {Beckmann}, {K{\"o}hler}, {Vanelle}, \& {Reimers}}]{2000A&A...358...77W}
{Wisotzki}, L., {Christlieb}, N., {Bade}, N., {et~al.} 2000, \aap, 358, 77.
\newblock \doarXiv{astro-ph/0004162}

\bibitem[{{Xu} {et~al.}(2012){Xu}, {Komossa}, {Zhou}, {Lu}, {Li}, {Grupe}, {Wang}, \& {Yuan}}]{2012AJ....143...83X}
{Xu}, D., {Komossa}, S., {Zhou}, H., {et~al.} 2012, \aj, 143, 83, \dodoi{10.1088/0004-6256/143/4/83}

\end{thebibliography}

\appendix

\section{Observing Log and Spectra}  
Table \ref{tab:Table of Targets} provides our observing log.  Signal-to-noise ratios were computed over 0.01 \textmu m ranges.  They were found in the centers of the J, H, and K bands for most objects except in cases were the central 0.01 \textmu m overlapped with an emission line.  In the table, ``combined" indicates that the final spectrum of the object was generated using data from multiple nights from the beginning of the reduction process.  ``Combined wa" 
indicates that the spectra for the object were combined using a weighted average after data from individual nights were reduced separately. In these cases, spectra were weighted by the square of the signal-to-noise ratios.  The signal-to-noise ratios in the J band were used to combine the J band portions of the spectra, and so on for the H and K bands. 

\begin{longtable*}{cccccccc}

  \tablewidth{0pt}
\tablecaption{Observing Log\label{tab:Table of Targets}}
\tablehead{
\colhead{Object}				&
\colhead{z}        &
\colhead{SNR J Band}	&
\colhead{SNR H Band}		&
\colhead{SNR J Band}				&
\colhead{Date Observed}			&
\colhead{Exposure Time (s)}		&
\colhead{Airmass}				
}
\startdata

PG 0003+158 & 0.451 & 42.69 & 86.60 & 64.08 & 08/27/21 & 6000 & 1.079\\
PG 0003+199 & 0.026 & 76.63 & 46.09 & 34.39 & 11/18/19 & 2400 & 3.895\\
PG 0007+106 & 0.089 & 67.08 & 71.75 & 120.47 & 12/06/19 & 6000 & 1.080\\
PG 0026+129 & 0.142 & 26.20 & 52.82 & 71.12 & 08/27/21 & 6000 & 1.068\\
PG 0043+039 & 0.385 & 30.67 & 8.14 & 8.63 & 11/18/2018 & 2880 & 1.983\\
PG 0043+039 & 0.385 & 13.82 & 16.28 & 16.45 & 08/27/2021 & 1200 & 1.373\\
PG 0043+039 & 0.385 & 14.18 & 14.79 & 13.72 & 08/28/2021  & 7200 & 2.651\\
PG 0043+039 & 0.385 & 29.27 & 25.44 & 23.21 & combined wa & combined wa & combined wa\\
PG 0049+171 & 0.064 & 39.51 & 45.01 & 22.50 & 09/19/21 & 2400 & 1.431\\
PG 0050+124 & 0.059 & 32.13 & 35.11 & 75.36 & 11/08/2019 & 3600 & 2.083\\
PG 0050+124 & 0.059 & 58.60 & 55.42 & 51.84 & 11/18/2019 & 3600 & 1.638\\
PG 0050+124 & 0.059 & 68.05 & 59.63 & 81.98 & combined wa & combined wa & combined wa\\
PG 0052+251 & 0.154 & 35.85 & 34.63 & 52.43 & 09/26/21 & 4800 & 1.214\\
PG 0157+001 & 0.163 & 52.34 & 86.94 & 73.25 & 12/06/19 & 2400 & 1.411\\
PG 0804+761 & 0.100	& 96.12 & 121.20 & 109.78 & 11/02/20 & 4800 & 1.886\\
PG 0838+770 & 0.131 & 26.01 & 36.17 & 48.38 & 11/02/20 & 2400 & 1.397\\
PG 0844+349 & 0.064 & 83.21 & 61.49 & 48.58 & 02/25/19 & 2400 & 1.634\\
PG 0921+525 & 0.035 & 83.18 & 73.70 & 73.55 & 02/25/19 & 3600 & 1.597\\
PG 0923+201 & 0.192 & 25.17 & 41.08 & 28.10 & 01/17/19 & 2400 & 1.056\\
PG 0923+129 & 0.029	& 88.05 & 79.55 & 25.73 & 01/17/19 & 1920 & 1.065\\
PG 0934+013 & 0.050	& 20.89 & 32.46 & 31.61 & 01/17/19 & 2400 & 1.191\\
PG 0947+396 & 0.206 & 30.33 & 57.37 & 79.22 & 12/11/22 & 7200 & 1.140\\
PG 0953+414 & 0.234	& 22.24 & 36.55 & 46.67 & 11/20/20 & 3600 & 1.153\\
PG 1001+054 & 0.161 & 43.86 & 68.93 & 57.05 & 01/16/22 & 6000 & 1.296\\
PG 1004+130	& 0.240 & 33.21 & 42.25 & 46.75 & 11/18/18 & 1200 & 1.185\\
PG 1011-040 & 0.058	& 23.17 & 33.49 & 58.40 & 11/18/18 & 960 & 1.372\\
PG 1012+008	& 0.187 & 27.60 & 23.09 & 33.17 & 11/18/18 & 1200 & 1.245\\
PG 1022+519 & 0.045 & 31.72 & 20.06 & 24.71 & 01/17/19 & 2400 & 1.122\\
PG 1048+342 & 0.167 & 17.00 & 28.60 & 29.52 & 03/19/22 & 10800 & 1.048\\
PG 1048-090 & 0.345 & 14.91 & 15.22 & 24.98 & 11/18/2019 & 3600 & 1.777\\
PG 1048-090 & 0.345 & 8.35 & 12.21 & 20.38 & 02/12/2022 & 7200 & 1.379\\
PG 1048-090 & 0.345 & 20.95 & 20.11 & 23.12 & combined & 10800 & combined\\
PG 1049-005 & 0.359 & 40.98 & 21.26 & 12.10 & 01/17/2019 & 2400 & 1.451\\
PG 1049-005 & 0.359 & 68.50 & 55.43 & 81.67 & 02/15/2022 & 6000 & 1.225\\
PG 1049-005 & 0.359 & 64.05 & 64.37 & 90.27 & combined wa & combined wa & combined wa\\
PG 1100+772 & 0.311 & 12.0 & 12.45 & 20.28 & 01/17/2019 & 1920 & 1.496\\
PG 1100+772 & 0.311 & 29.31 & 62.65 & 42.77 & 02/25/2022 & 5400 & 1.414\\
PG 1100+772 & 0.311 & 25.30 & 66.84 & 46.65 & combined wa & combined wa & combined wa\\
PG 1103-006 & 0.423 & 28.70 & 32.95 & 64.45 & 01/09/23 & 10800 & 1.951\\ 
PG 1114+445 & 0.144 & 33.99 & 71.35 & 76.47 & 01/09/23 & 8400 & 1.021\\
PG 1115+407 & 0.154 & 35.45 & 40.15 & 120.32 & 01/18/22 & 6000 & 1.101\\
PG 1116+215 & 0.177 & 28.39 & 54.97 & 123.28 & 11/18/18 & 960 & 1.113\\
PG 1119+120	& 0.050 & 91.66 & 117.59 & 108.55 & 12/14/19 & 3600 & 1.230\\
PG 1121+422	& 0.225 & 30.99 & 41.33 & 21.96 & 01/25/19 & 3600 & 1.153\\
PG 1126-041	& 0.062 & 116.00 & 80.99 & 239.34 & 02/04/23 & 8400 & 1.473\\
PG 1149-110	& 0.049 & 43.27 & 20.38 & 38.07 & 01/25/19 & 2400 & 1.476\\
PG 1151+117	& 0.176 & 26.37 & 20.51 & 44.20 & 01/25/19 & 3600 & 1.078\\
PG 1202+281	& 0.165 & 27.27 & 31.72 & 30.21 & 01/25/19 & 2400 & 1.016\\
PG 1211+143	& 0.081 & 64.29 & 61.99 & 71.40 & 06/15/19 & 3600 & 1.097\\
PG 1216+069	& 0.331 & 34.81 & 26.36 & 29.00 & 06/14/19 & 3600 & 1.332\\
PG 1226+023	& 0.158 & 55.79 & 246.58 & 85.25 & 12/14/19 & 4800 & 1.334\\
PG 1229+204	& 0.063 & 104.95 & 63.37 & 41.39 & 04/23/21 & 6000 & 1.311\\		
PG 1244+026	& 0.048 & 55.14 & 33.80 & 92.00 & 02/04/23 & 8400 & 1.160\\
PG 1259+593 & 0.478 & 46.15 & 57.56 & 39.81 & 03/19/22 & 4800 & 1.276\\
PG 1302-102 & 0.278 & 23.41 & 28.62 & 41.58 & 04/17/22 & 2400 & 1.673\\
PG 1307+085 & 0.155 & 14.21 & 24.57 & 34.77 & 02/19/2019 & 2400 & 1.122\\
PG 1307+085 & 0.155 & 27.96 & 24.07 & 26.04 & 04/17/2022 & 7200 & 1.585\\
PG 1307+085 & 0.155 & 33.09 & 26.84 & 60.94 & combined wa & combined wa & combined wa\\
PG 1309+355	& 0.183 & 45.82 & 56.84 & 30.81 & 07/09/19 & 2700 & 1.060\\
PG 1310-108	& 0.034 & 21.42 & 27.79 & 36.04 & 02/19/19 & 2400 & 1.670\\
PG 1322+659 & 0.168 & 22.06 & 34.77 & 134.45 & 01/18/22 & 6000 & 1.255\\
PG 1341+258	& 0.087 & 49.52 & 46.70 & 53.34 & 04/19/19 & 3600 & 1.021\\
PG 1351+236	& 0.055 & 46.35 & 51.99 & 73.19 & 04/23/21 & 4800 & 1.165\\	
PG 1351+640	& 0.088 & 65.28 & 88.66 & 89.98 & 04/19/19 & 3600 & 1.244\\
PG 1352+183 & 0.152 & 20.41 & 23.10 & 28.60 & 06/30/22 & 8700 & 1.037\\
PG 1354+213 & 0.300 & 20.70 & 36.73 & 48.88 & 02/11/23 & 10800 & 1.756\\ 
PG 1402+261	& 0.164 & 21.91 & 62.73 & 34.47 & 07/09/19 & 3000 & 1.135\\
PG 1404+226 & 0.098 & 19.29 & 13.71 & 14.37 & 07/09/2019 & 3600 & 1.396\\
PG 1404+226 & 0.098 & 16.86 & 30.80 & 20.44 & 07/10/2022 & 6000 & 1.478\\
PG 1404+226 & 0.098 & 17.91 & 31.20 & 22.78 & combined wa & 9600 & combined wa\\
PG 1411+442	& 0.090 & 79.62 & 62.93 & 81.10 & 07/09/19 & 1800 & 1.593\\	
PG 1415+451	& 0.114 & 44.41 & 84.62 & 67.96 & 07/14/19 & 3600 & 1.044\\
PG 1416-129 & 0.129 & 10.70 & 9.84 & 12.13 & 07/14/2019 & 3600 & 1.78\\
PG 1416-129 & 0.129 & 16.91 & 15.18 & 26.68 & 07/10/2022 & 7200 & 1.480\\
PG 1416-129 & 0.129 & 13.46 & 12.49 & 30.33 & 02/11/2023 & 8400 & 1.508\\
PG 1416-129 & 0.129 & 15.16 & 9.97 & 8.58 & 07/01/2023 & 6000 & 1.899\\
PG 1416-129 & 0.129 & 15.92 & 23.13 & 50.19 & 07/04/2023 & 7200 & 1.444\\
PG 1416-129 & 0.129 & 18.82 & 22.77 & 38.43 & combined wa & combined wa & combined wa\\
PG 1425+267 & 0.364 & 32.65 & 51.29 & 48.0 & 04/12/23 & 4200 & 1.013\\
PG 1426+015	& 0.087 & 82.00 & 73.88 & 101.99 & 05/17/21 & 6000 & 1.674\\	
PG 1427+480 & 0.220 & 25.40 & 32.46 & 44.04 & 05/15/22 & 7200 & 1.093\\
PG 1435-067	& 0.126 & 20.47 & 21.32 & 32.49 & 06/17/21 & 4800 & 1.393\\	
PG 1440+356	& 0.079 & 158.38 & 98.48 & 151.61 & 05/05/21 & 6000 & 1.120\\		
PG 1444+407 & 0.267 & 58.49 & 43.13 & 59.32 & 05/15/22 & 6000 & 1.400\\
PG 1448+273	& 0.065 & 51.49 & 40.52 & 90.02 & 05/08/22 & 6000 & 1.348\\
PG 1501+106	& 0.036 & 58.03 & 88.94 & 156.83 & 06/17/21 & 4800 & 1.081\\
PG 1512+370 & 0.371 & 24.30 & 28.24 & 21.05 & 07/04/23 & 7200 & 1.077\\
PG 1519+226	& 0.137 & 50.10 & 85.81 & 63.42 & 05/24/22 & 4800 & 1.072\\	
PG 1534+580 & 0.030 & 90.48 & 32.21 & 97.03 & 05/11/21 & 6000\\	
PG 1535+547	& 0.039 & 58.03 & 88.94 & 156.83 & 06/17/21 & 2400 & 1.105\\
PG 1543+489 & 0.401 & 11.75 & 19.56 & 23.56 & 07/04/2020 & 3600 & 1.131\\
PG 1543+489 & 0.401 & 9.38 & 7.78 & 10.54 & 10/18/2020 & 3600 & 1.494\\
PG 1543+489 & 0.401 & 19.76 & 22.47 & 22.50 & 10/22/2021 & 3600 & 1.499\\
PG 1543+489 & 0.401 & 23.17 & 31.76 & 44.12 & combined & 10800 & combined\\
PG 1545+210	& 0.264 & 32.34 & 40.92 & 39.08 & 08/25/21 & 6000 & 1.160\\
PG 1552+085	& 0.119 & 33.14 & 49.06 & 49.00 & 04/12/23 & 6000 & 1.099\\
PG 1612+261	& 0.131 & 25.22 & 70.32 & 49.11 & 08/28/21 & 6000 & 1.076\\
PG 1613+658	& 0.129 & 23.77 & 82.09 & 75.18 & 06/30/19 & 4800 & 1.642\\
PG 1617+175	& 0.112 & 48.05 & 73.95 & 114.27 & 05/22/19 & 3600 & 1.037\\
PG 1626+554	& 0.133 & 25.88 & 46.11 & 94.20 & 05/22/2019 & 3600 & 1.111\\
PG 1626+554	& 0.133 & 11.87 & 15.43 & 20.10 & 10/18/2020 & 3600 & 1.653\\
PG 1626+554	& 0.133 & 28.85 & 43.96 & 97.70 & combined wa & combined wa & combined wa\\
PG 1700+518	& 0.292 & 43.47 & 67.38 & 91.70 & 05/22/19 & 1200 & 1.273\\
PG 1704+608	& 0.372 & 31.07 & 66.48 & 74.98 & 05/22/19 & 3600 & 1.196\\
PG 2112+059	& 0.466 & 43.05 & 58.80 & 54.16 & 10/22/21 & 15600 & 1.117\\		
PG 2130+099	& 0.063 & 96.45 & 67.56 & 27.41 & 07/04/20 & 3600 & 3.219\\
PG 2209+184	& 0.070 & 49.32 & 43.76 & 50.56 & 08/20/21 & 6000 & 1.175\\	
PG 2214+139	& 0.066 & 107.00 & 82.77 & 160.19 & 12/06/19 & 3000 & 1.054\\
PG 2233+134 & 0.326 & 12.11 & 16.31 & 17.44 & 08/20/2021 & 4800 & 2.865\\
PG 2233+134 & 0.326 & 21.29 & 23.39 & 25.88 & 09/19/2021 & 4800 & 1.155\\
PG 2233+134 & 0.326 & 25.17 & 35.49 & 35.20 & combined wa & combined wa & combined wa\\
PG 2251+113	& 0.326 & 69.66 & 70.90 & 56.87 & 08/25/21 & 7200 & 1.385\\	
PG 2304+042	& 0.042 & 72.58 & 85.63 & 29.07 & 08/11/19 & 2400 & 1.392\\
PG 2308+098	& 0.433 & 32.60 & 48.03 & 25.86 & 08/11/19 & 3600 & 1.112

\end{longtable*}


\section{Spectra}

\begin{figure*}
\begin{center}
\includegraphics[width=1.00\linewidth]{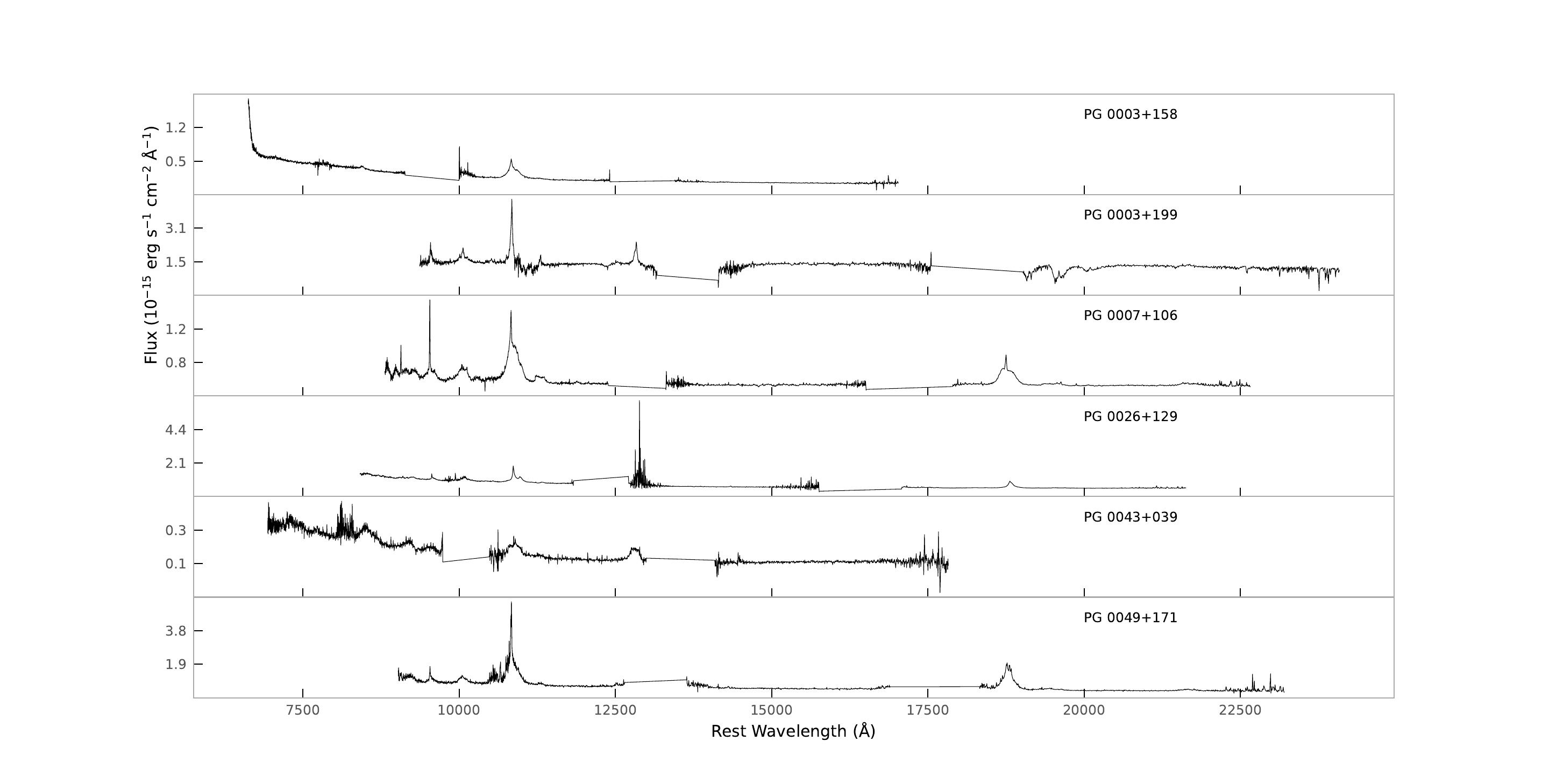}
\caption{\label{fig:A1}Spectra shifted to the rest frame}
\end{center}

\begin{center}
\includegraphics[width=1.00\linewidth]{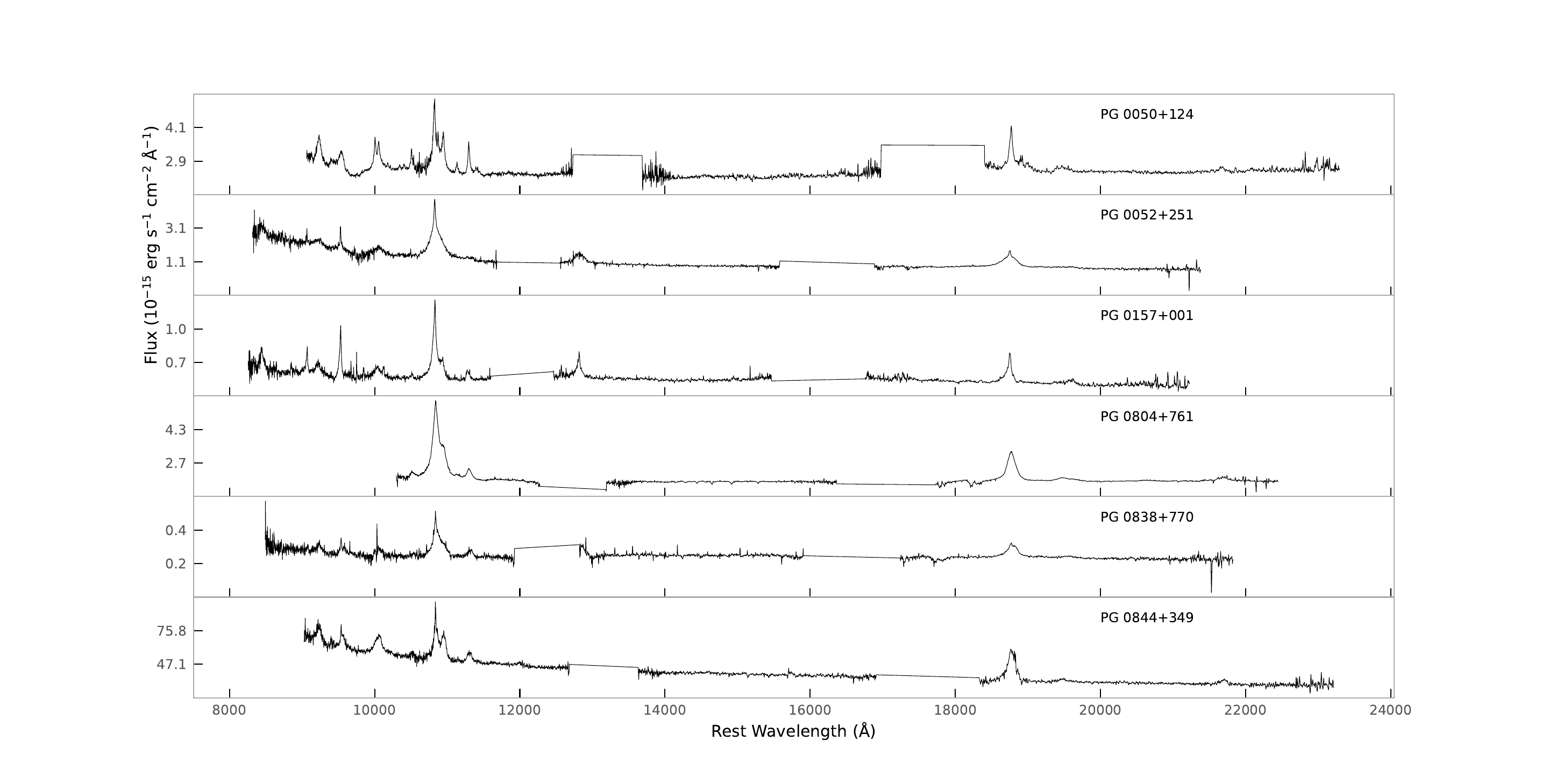}
\caption{\label{fig:A2}Spectra shifted to the rest frame}
\end{center}
\end{figure*}

\begin{figure}
\begin{center}
\includegraphics[width=1.00\linewidth]{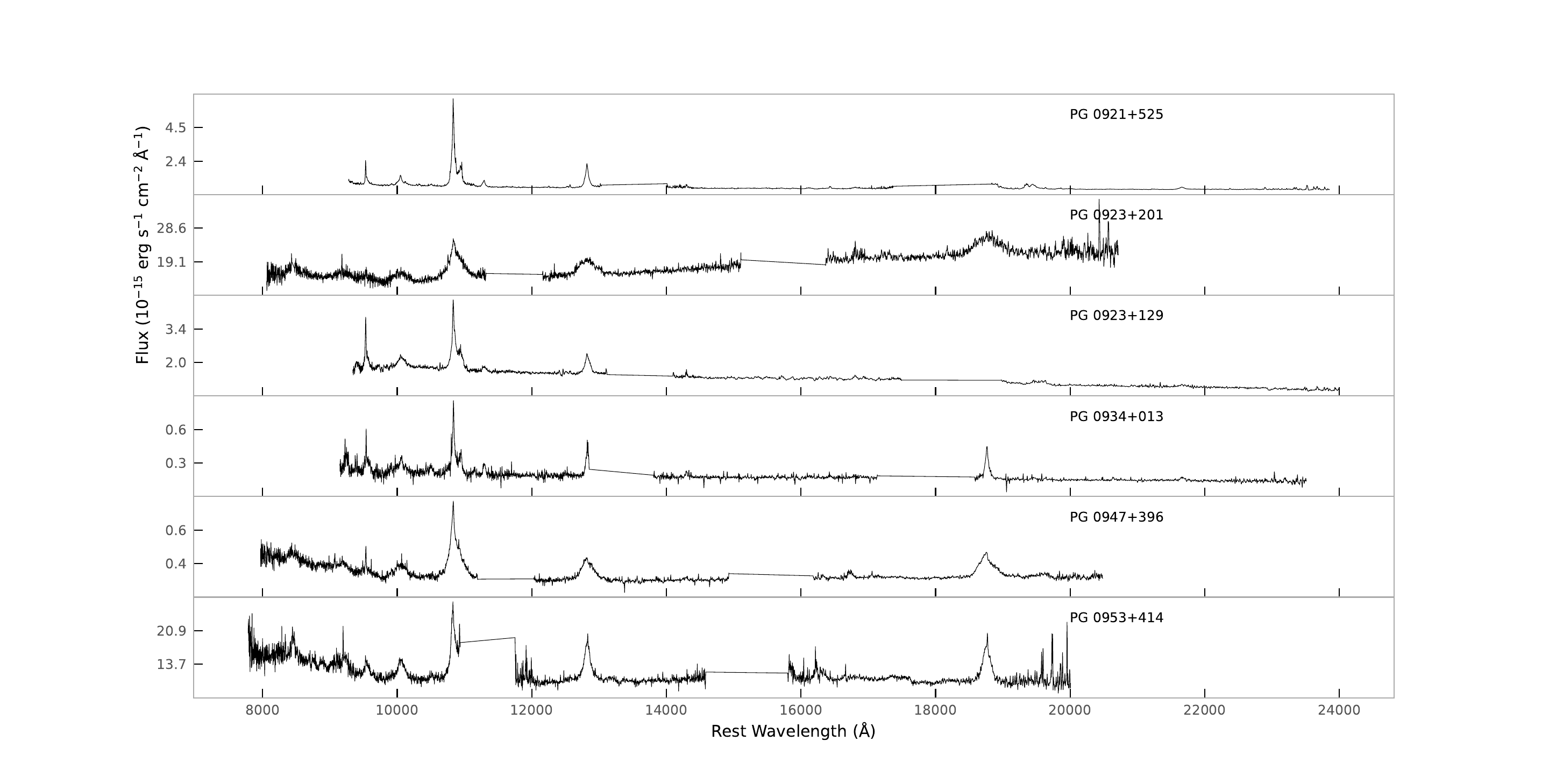}
\caption{\label{fig:A3}Spectra shifted to the rest frame}
\end{center}

\begin{center}
\includegraphics[width=1.00\linewidth]{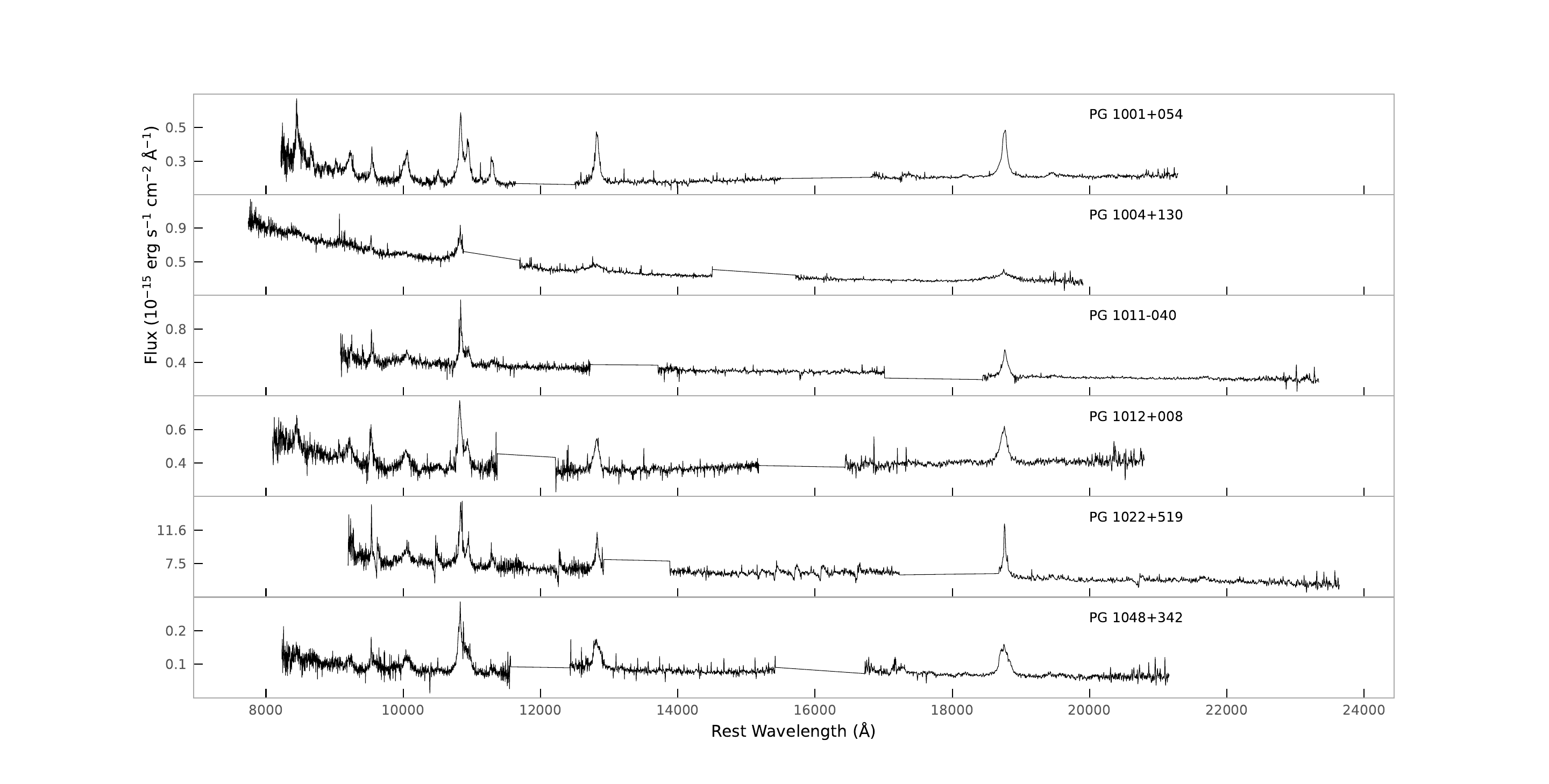}
\caption{\label{fig:A4}Spectra shifted to the rest frame}
\end{center}
\end{figure}

\begin{figure}
\begin{center}
\includegraphics[width=1.00\linewidth]{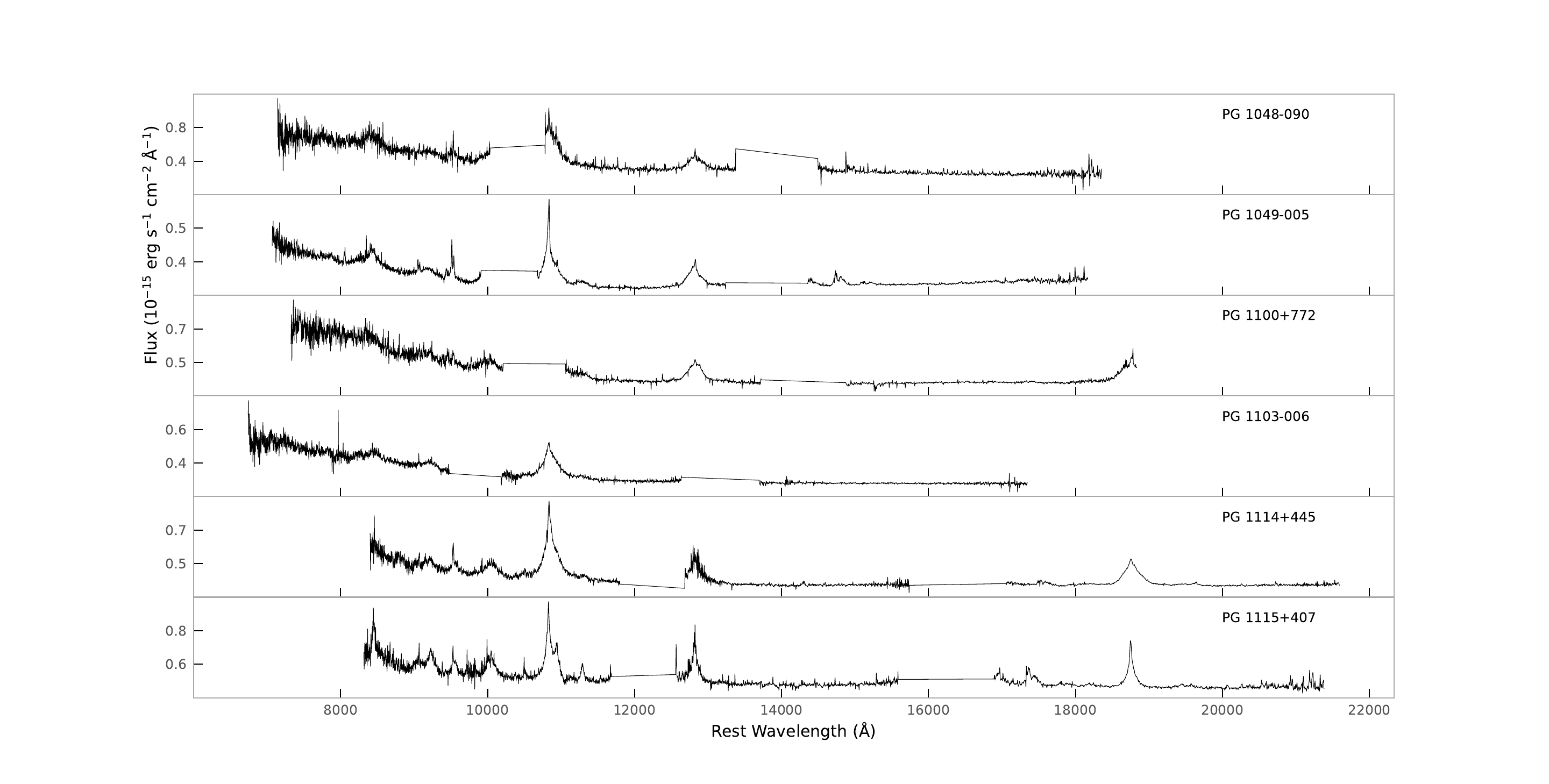}
\caption{\label{fig:A5}Spectra shifted to the rest frame}
\end{center}

\begin{center}
\includegraphics[width=1.00\linewidth]{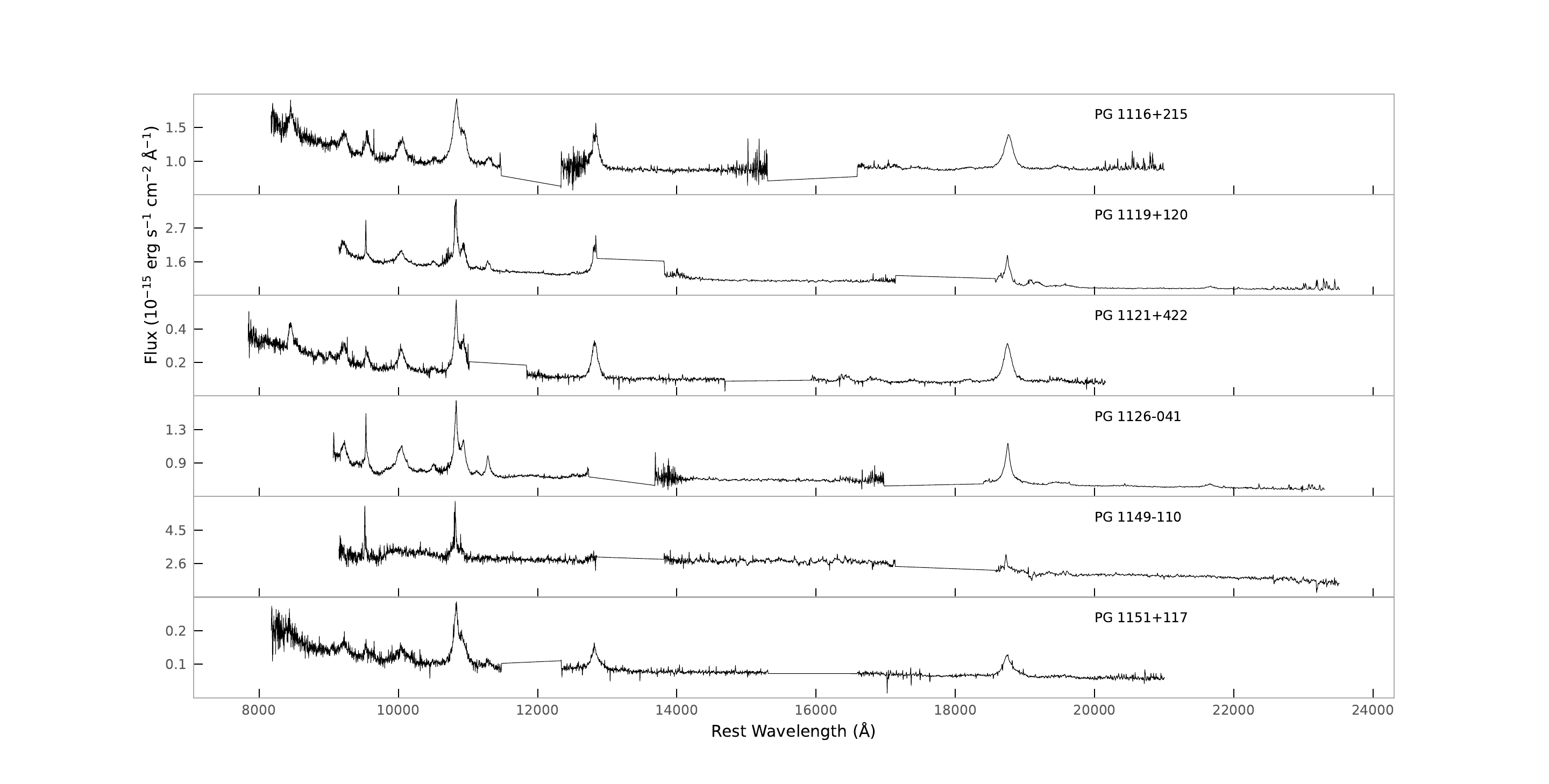}
\caption{\label{fig:A6}Spectra shifted to the rest frame}
\end{center}
\end{figure}

\begin{figure}
\begin{center}
\includegraphics[width=1.00\linewidth]{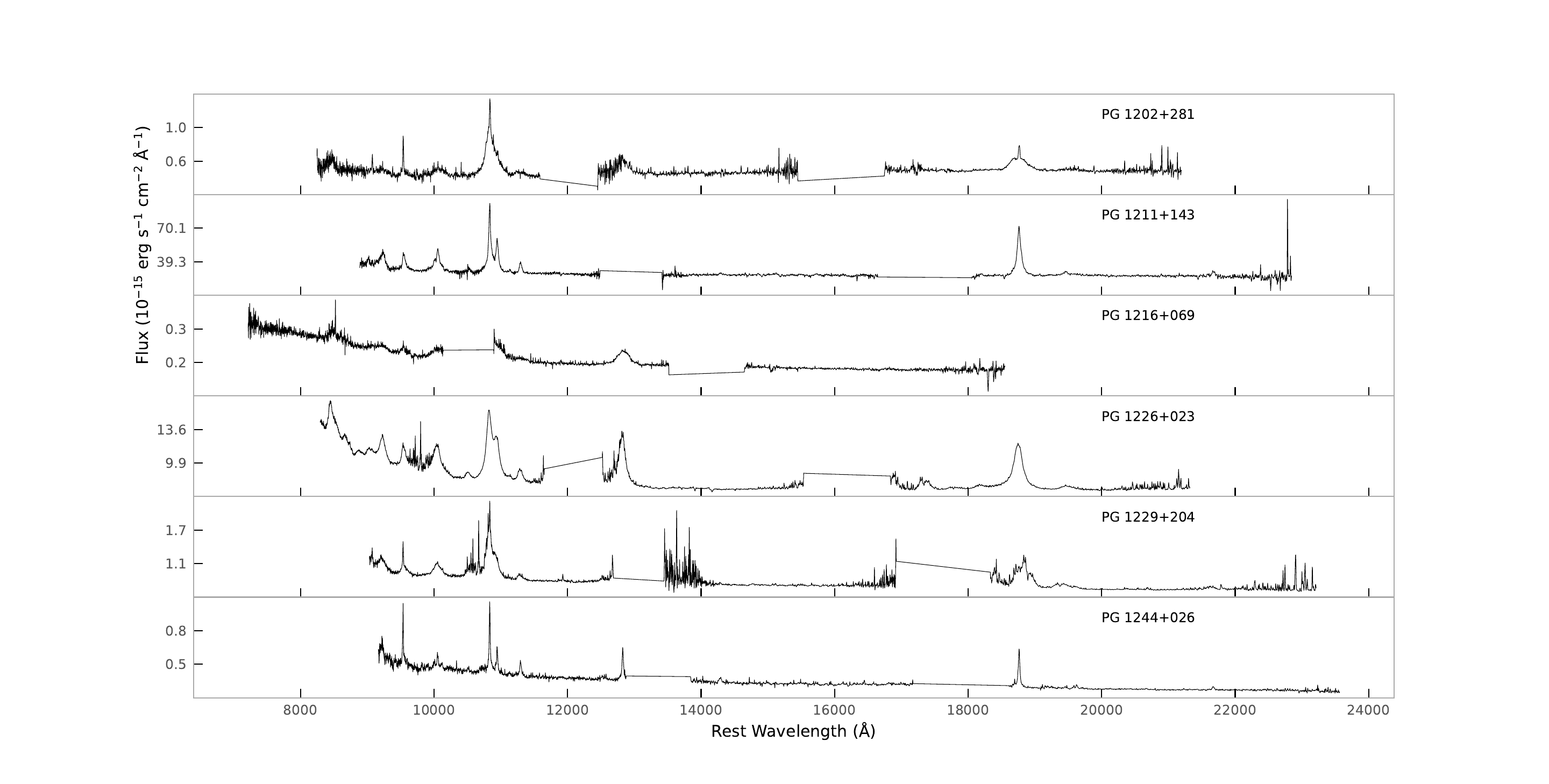}
\caption{\label{fig:A7}Spectra shifted to the rest frame}
\end{center}

\begin{center}
\includegraphics[width=1.00\linewidth]{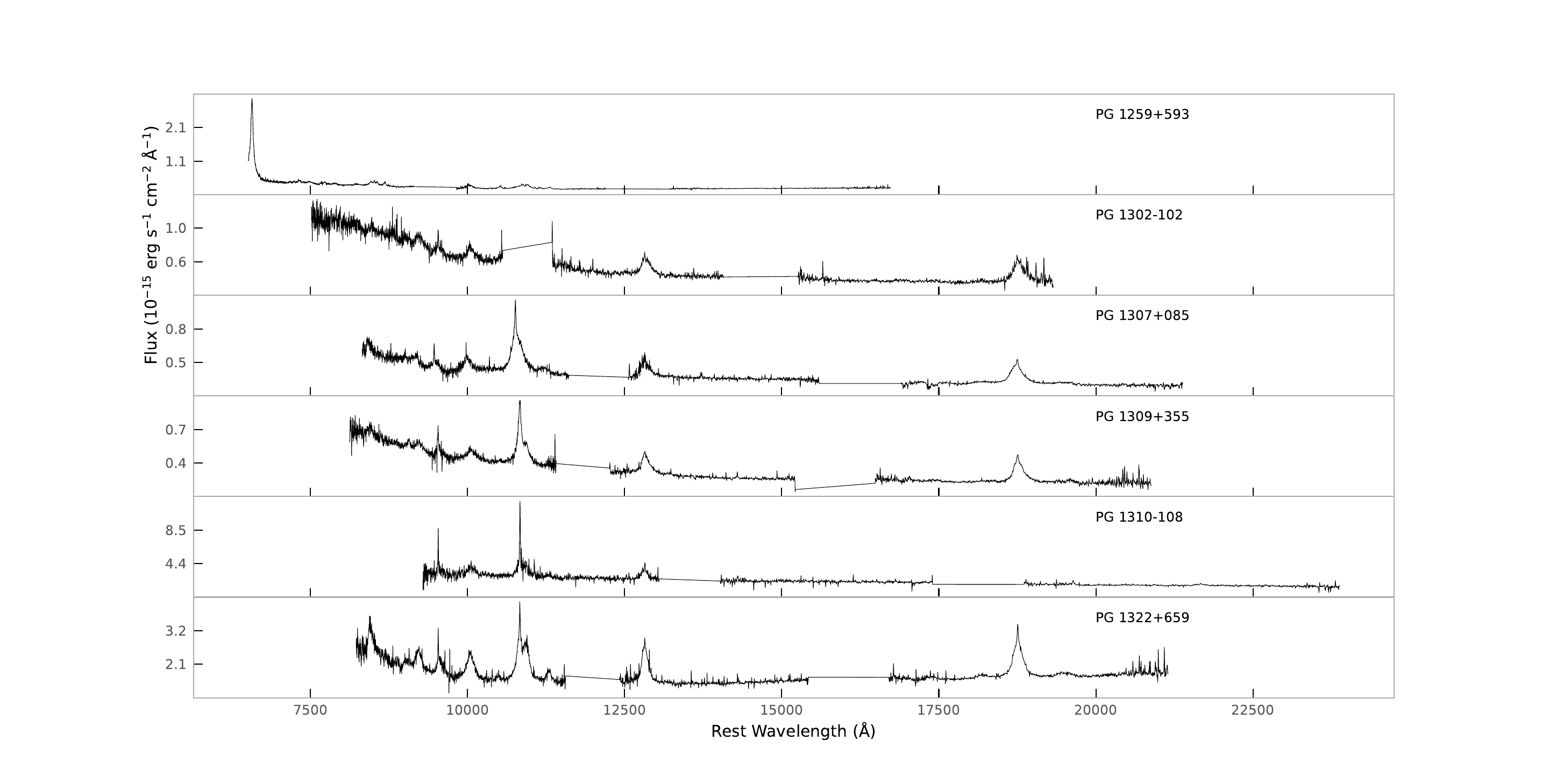}
\caption{\label{fig:A8}Spectra shifted to the rest frame}
\end{center}
\end{figure}

\begin{figure}
\begin{center}
\includegraphics[width=1.00\linewidth]{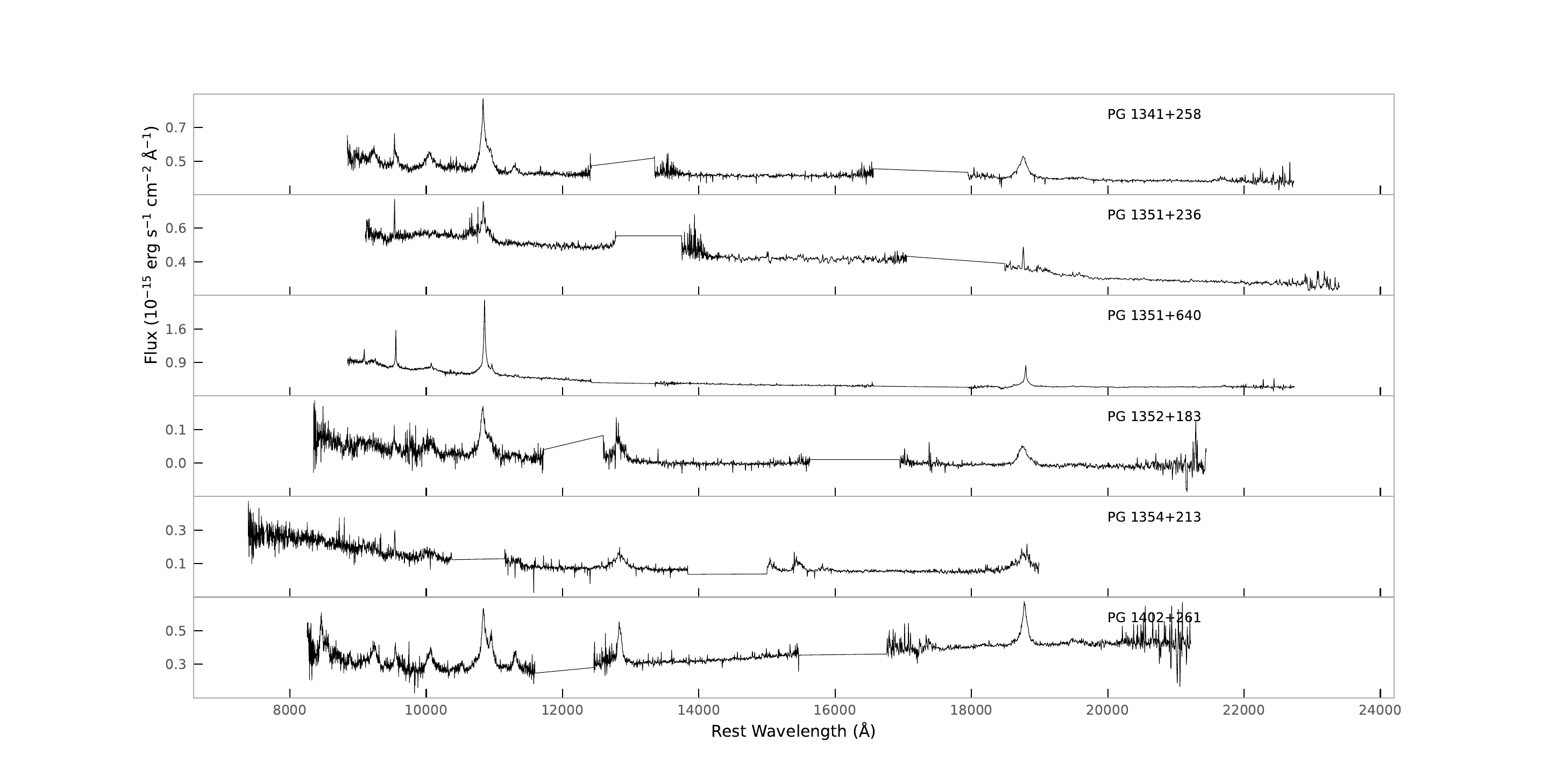}
\caption{\label{fig:A9}Spectra shifted to the rest frame}
\end{center}

\begin{center}
\includegraphics[width=1.00\linewidth]{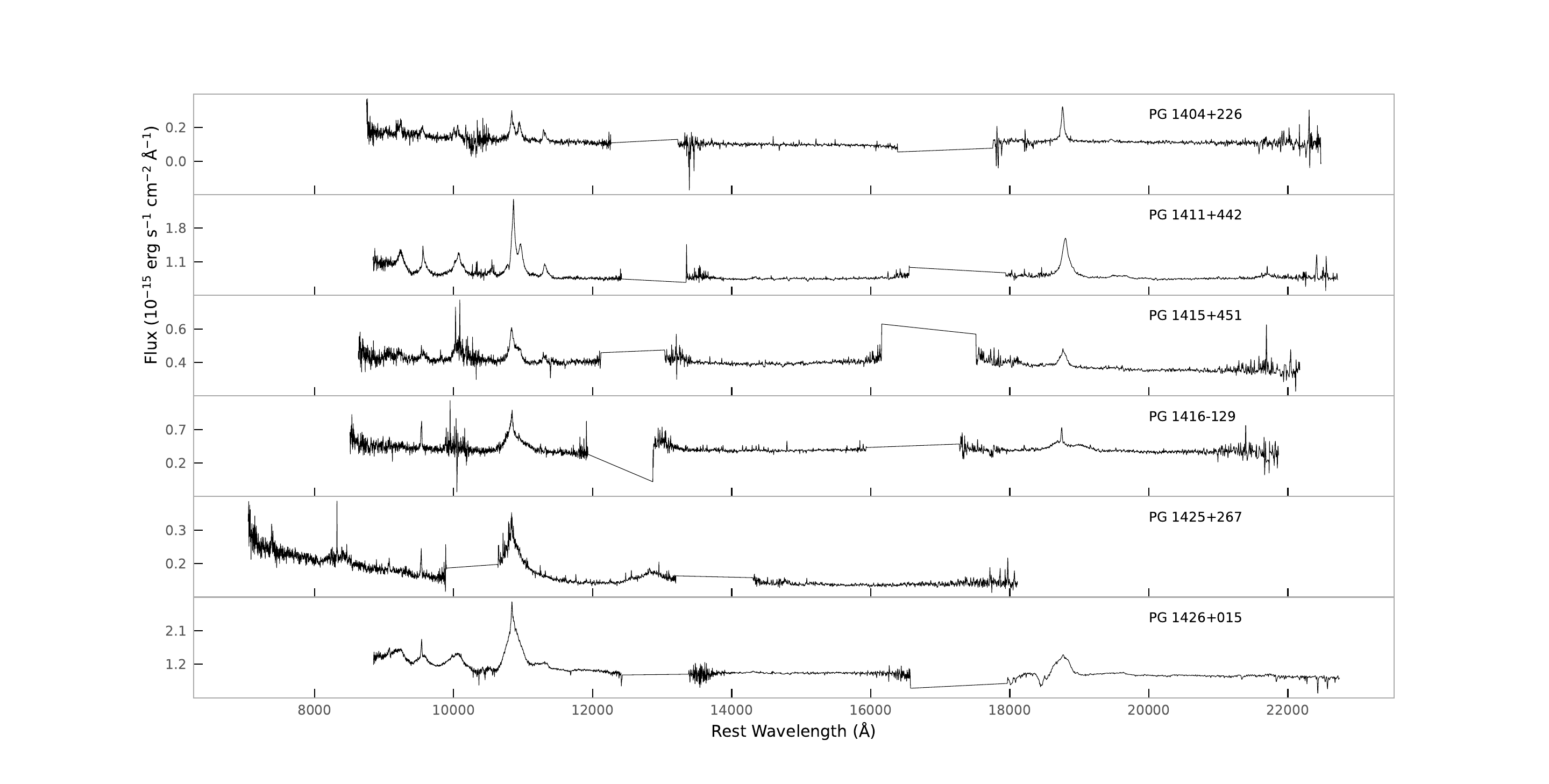}
\caption{\label{fig:A10}Spectra shifted to the rest frame}
\end{center}
\end{figure}

\begin{figure}
\begin{center}
\includegraphics[width=1.00\linewidth]{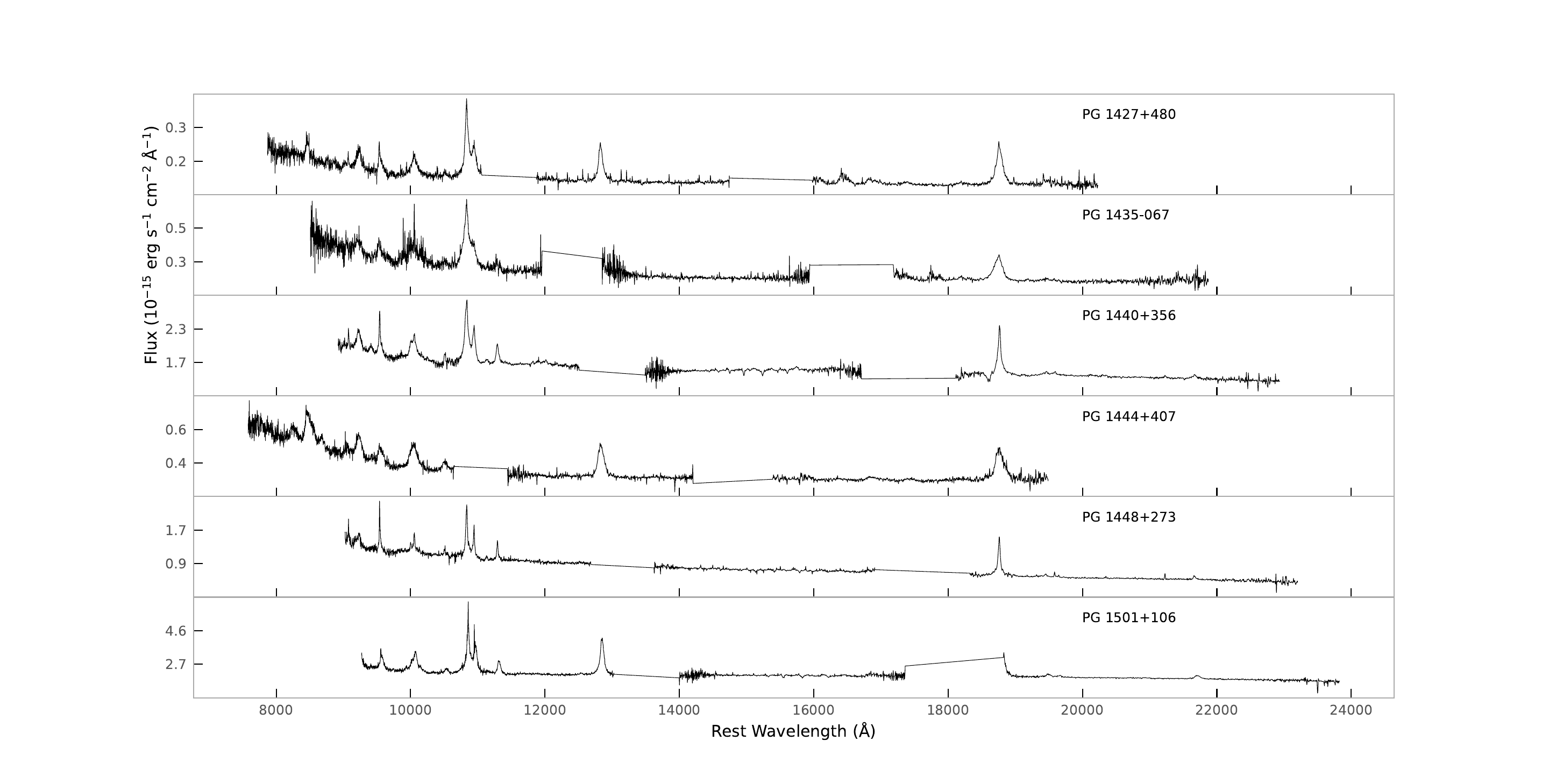}
\caption{\label{fig:A11}Spectra shifted to the rest frame}
\end{center}

\begin{center}
\includegraphics[width=1.00\linewidth]{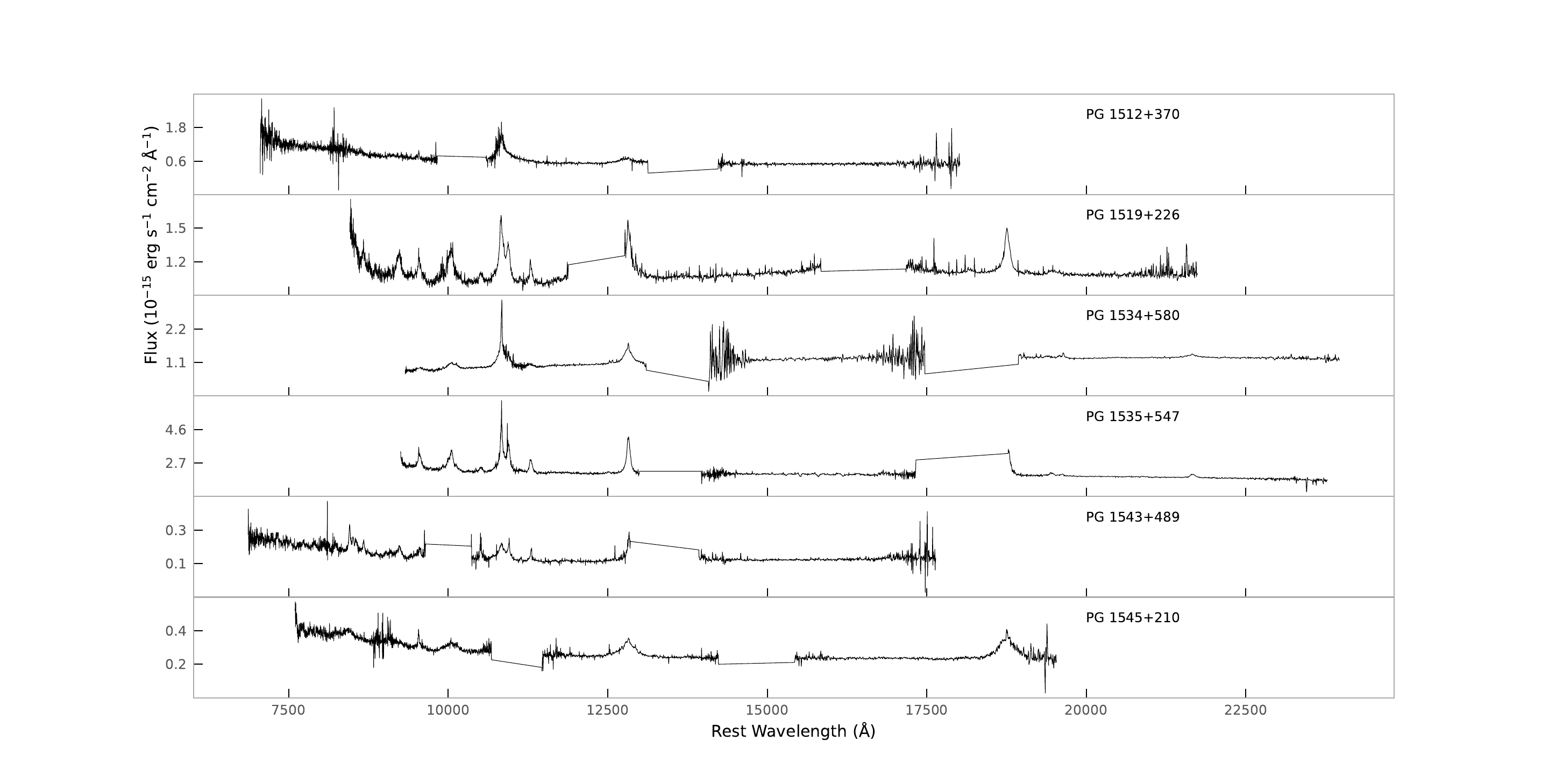}
\caption{\label{fig:A12}Spectra shifted to the rest frame}
\end{center}
\end{figure}

\begin{figure}
\begin{center}
\includegraphics[width=1.00\linewidth]{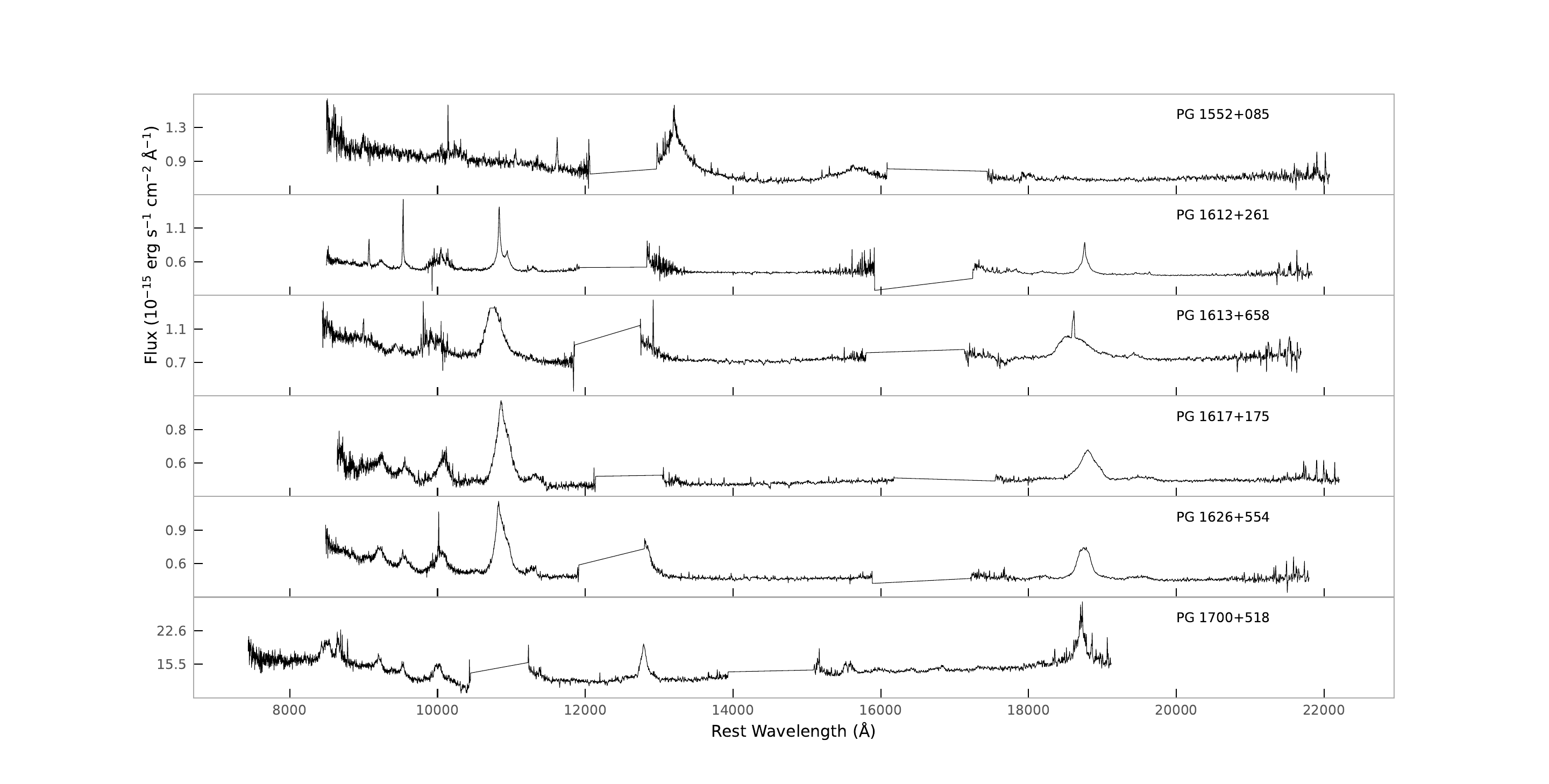}
\caption{\label{fig:A13}Spectra shifted to the rest frame}
\end{center}

\begin{center}
\includegraphics[width=1.00\linewidth]{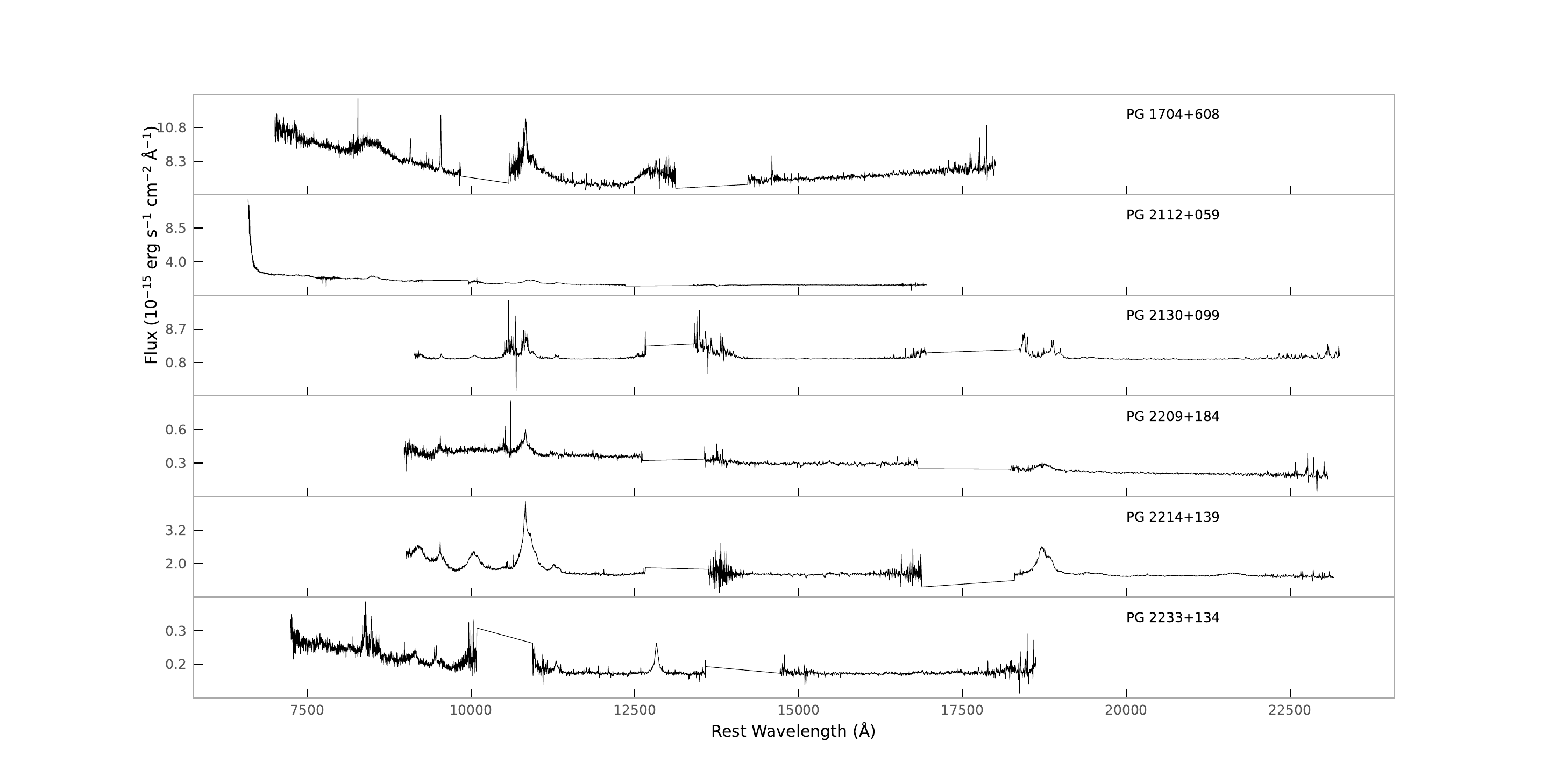}
\caption{\label{fig:A14}Spectra shifted to the rest frame}
\end{center}
\end{figure}

\begin{figure}
\begin{center}
\includegraphics[width=1.00\linewidth]{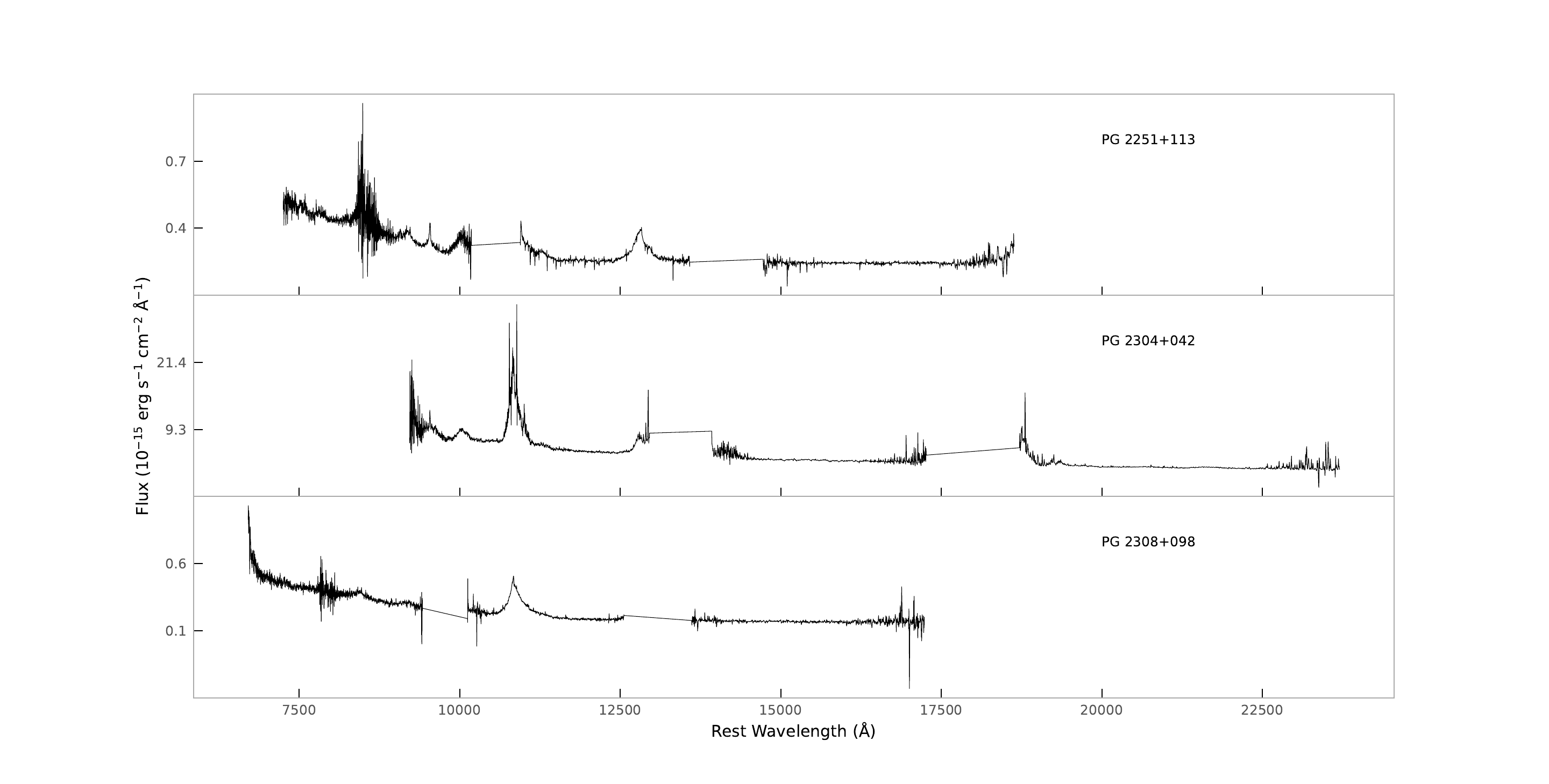}
\caption{\label{fig:A15}Spectra shifted to the rest frame}
\end{center}
\end{figure}

Figures \ref{fig:A1} through \ref{fig:A15} show our NIR spectra for the 87 quasars in the rest frame.



\newpage

\subsection{Composite Continuum and Line Measurements}

\begin{figure*}[!ht]
\centering
\includegraphics[width=1\linewidth]{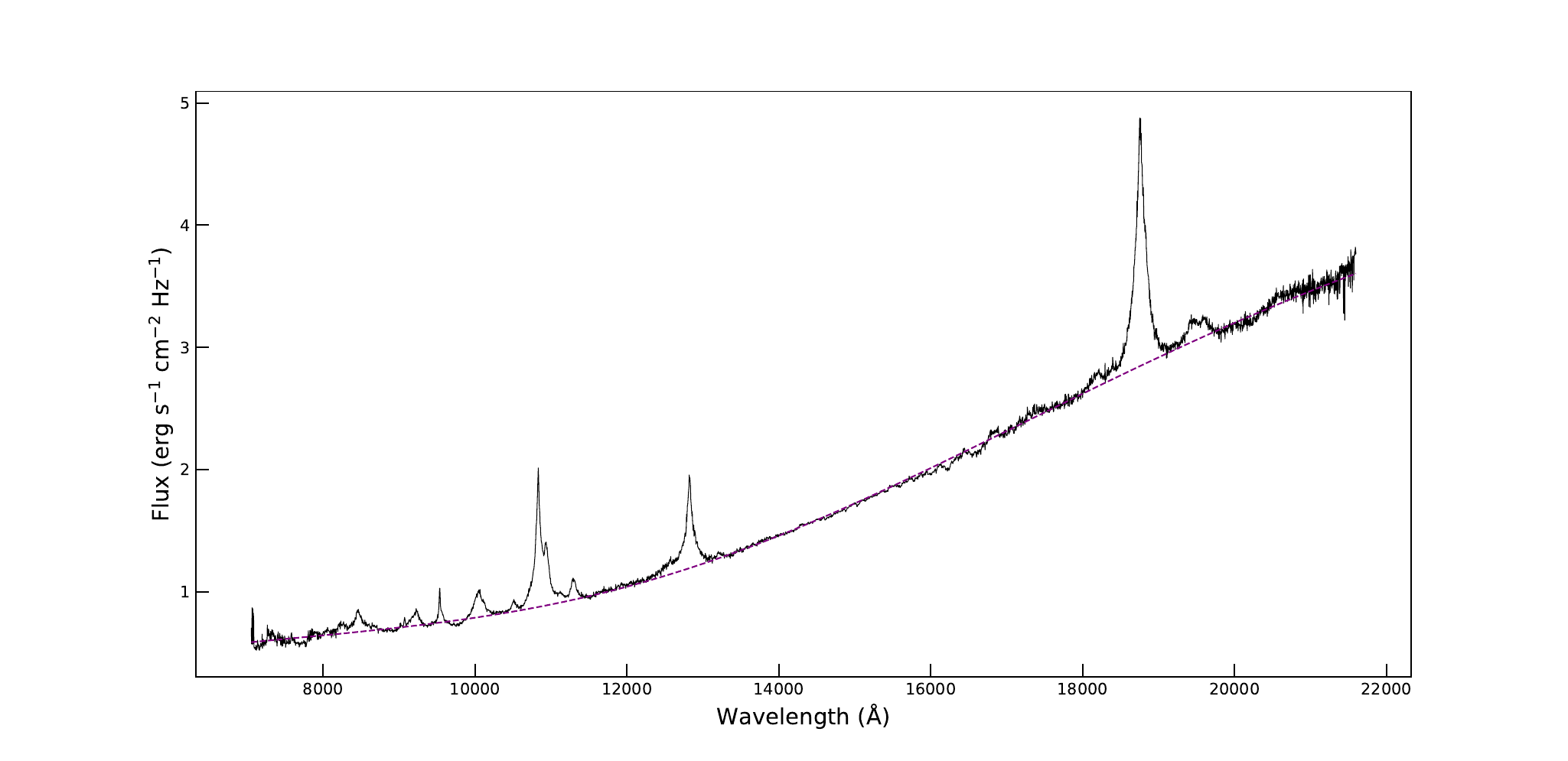}
\caption{\label{bbpowerlaw}Composite spectrum, fit with a blackbody and power law (magenta dashed line).}
\end{figure*}

Figure \ref{bbpowerlaw} is a combined power law and blackbody fit to the near-infrared spectrum, one conventional approach to modeling the NIR continuum. This is likely too simple but provides a characterization of the continuum. Other functional forms, such as broken power laws or polynomials may also provide good fits.  The flux is in arbitrary units of F$_{\nu}$ in order to show more clearly the blackbody plus power-law fit.
The equation used for the fit is

\begin{equation}
a \Big(\frac{2hc^{2}}{\lambda^{5} e^{hc / \lambda kT} - 1}\Big) + b\lambda^{\alpha}
\end{equation}

Here, \textalpha{} = 0.680, T = 985.6 K, a = 7.97 x 10$^{-10}$, and b = 9006.  Since it is difficult to fit a power law to the spectrum in the presence of multiple emission lines, only some wavelength regions are included in the fit.  Regions below 8080 \AA, from 8750-8950 \AA, from 9300-9430 \AA, 9700-9800 \AA, 10200-10400 \AA, 11400-12260 \AA, 13100-18450 \AA, and 19700-20400 \AA{} are included in the fit.  Before the fit, F$_{\lambda}$ was converted to F$_{\nu}$ and this converted flux was scaled by dividing by the mean F$_{\nu}$ flux in a region free of emission lines from 11600-11700 \AA. \par
\begin{figure*}[!ht]
\centering
\includegraphics[width=1\linewidth]{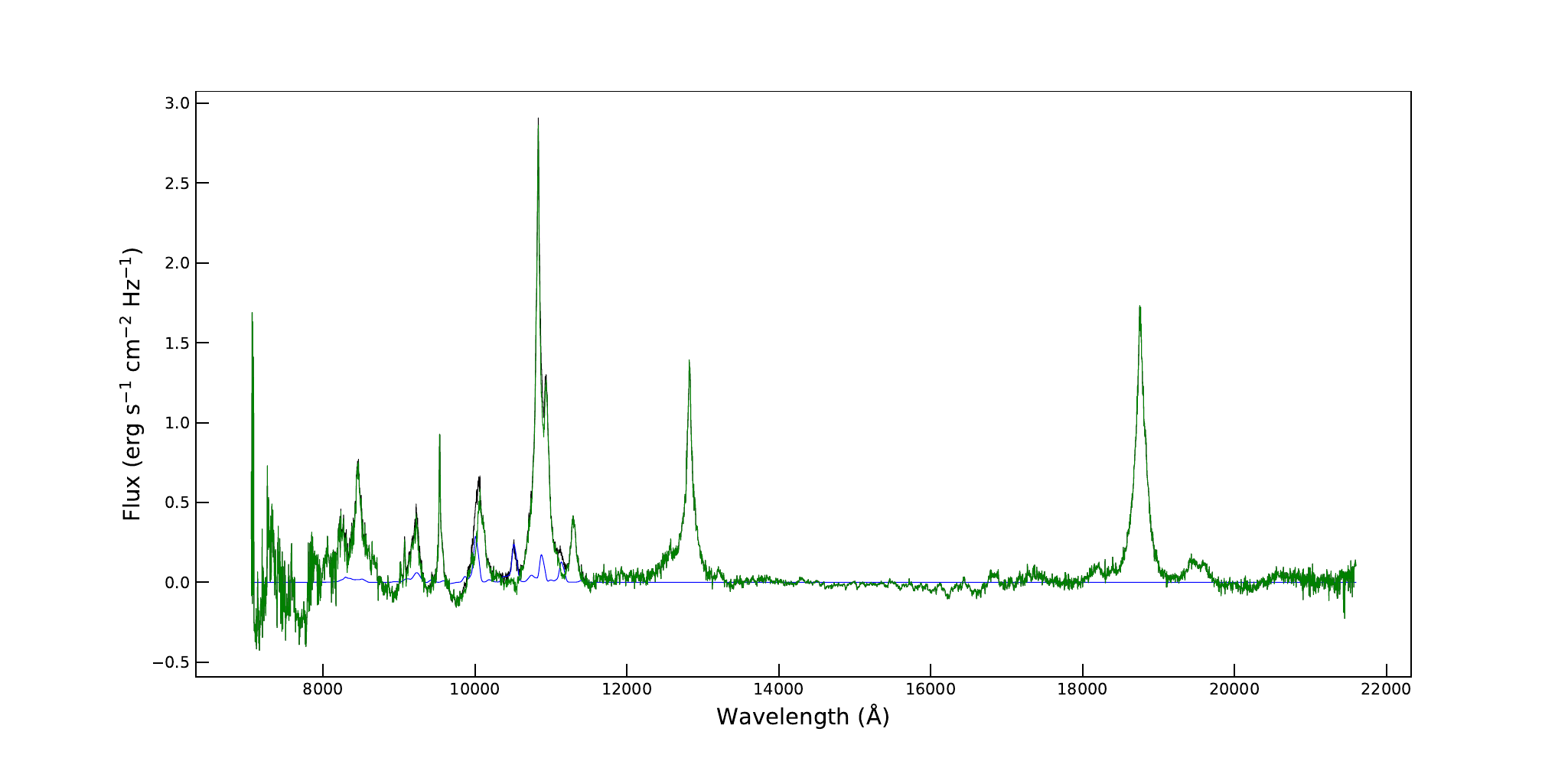}
\caption{\label{feiifit}Continuum-subtracted composite spectrum in green, along with a Fe II template in blue given by \citet{2012ApJ...751....7G}.}
\end{figure*}

In order to measure individual emission lines in the composite, we first fit an Fe II template.  Figure \ref{feiifit} shows the power-law-blackbody-subtracted composite along with an Fe II template from \citet{2012ApJ...751....7G} convolved with a Gaussian function with \textsigma{} = 20 \AA\ and scaled by a factor of 0.009. \par

\begin{figure*}[!ht]
\centering
\includegraphics[scale=0.5]{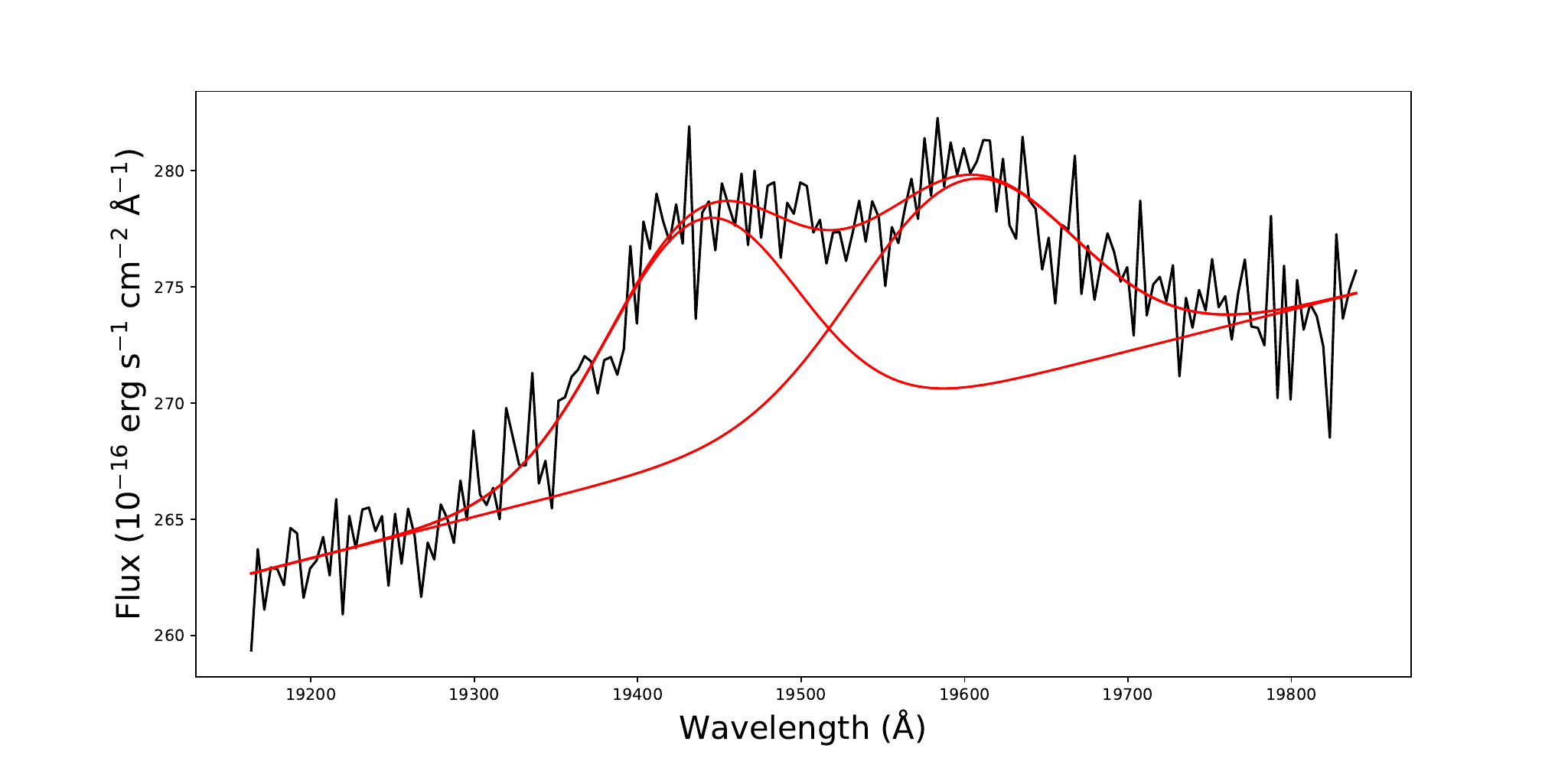}
\caption{\label{fig:line1}Composite spectrum, fit to Brackett Delta line (19446 \AA) blended with H$_2$\ at 19570\AA.}

\centering
\includegraphics[scale=0.5]{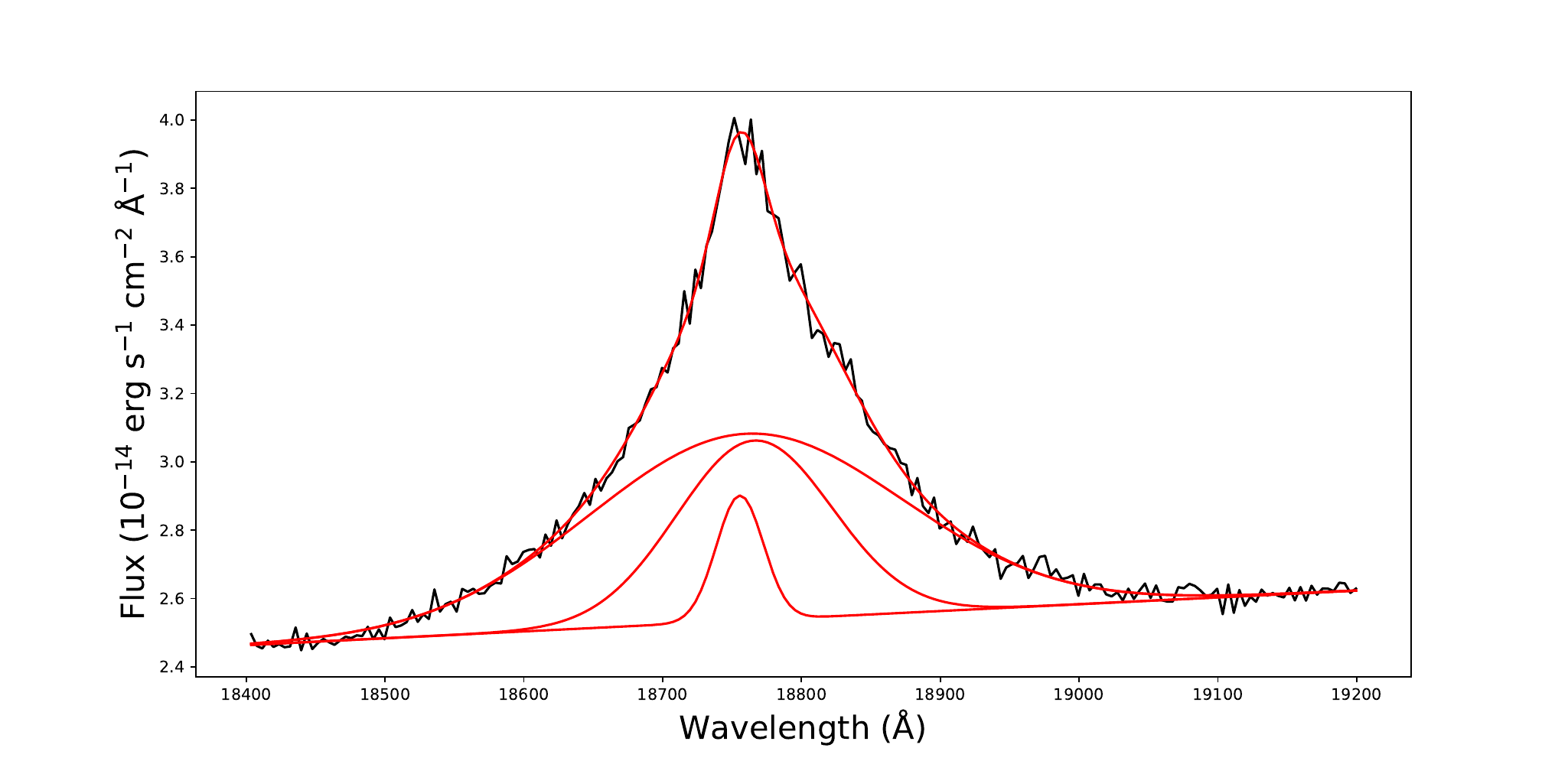}
\caption{\label{fig:line2}Composite spectrum, fit to Paschen Alpha line.}
\end{figure*}

\begin{figure*}[!ht]
\includegraphics[scale=0.5]{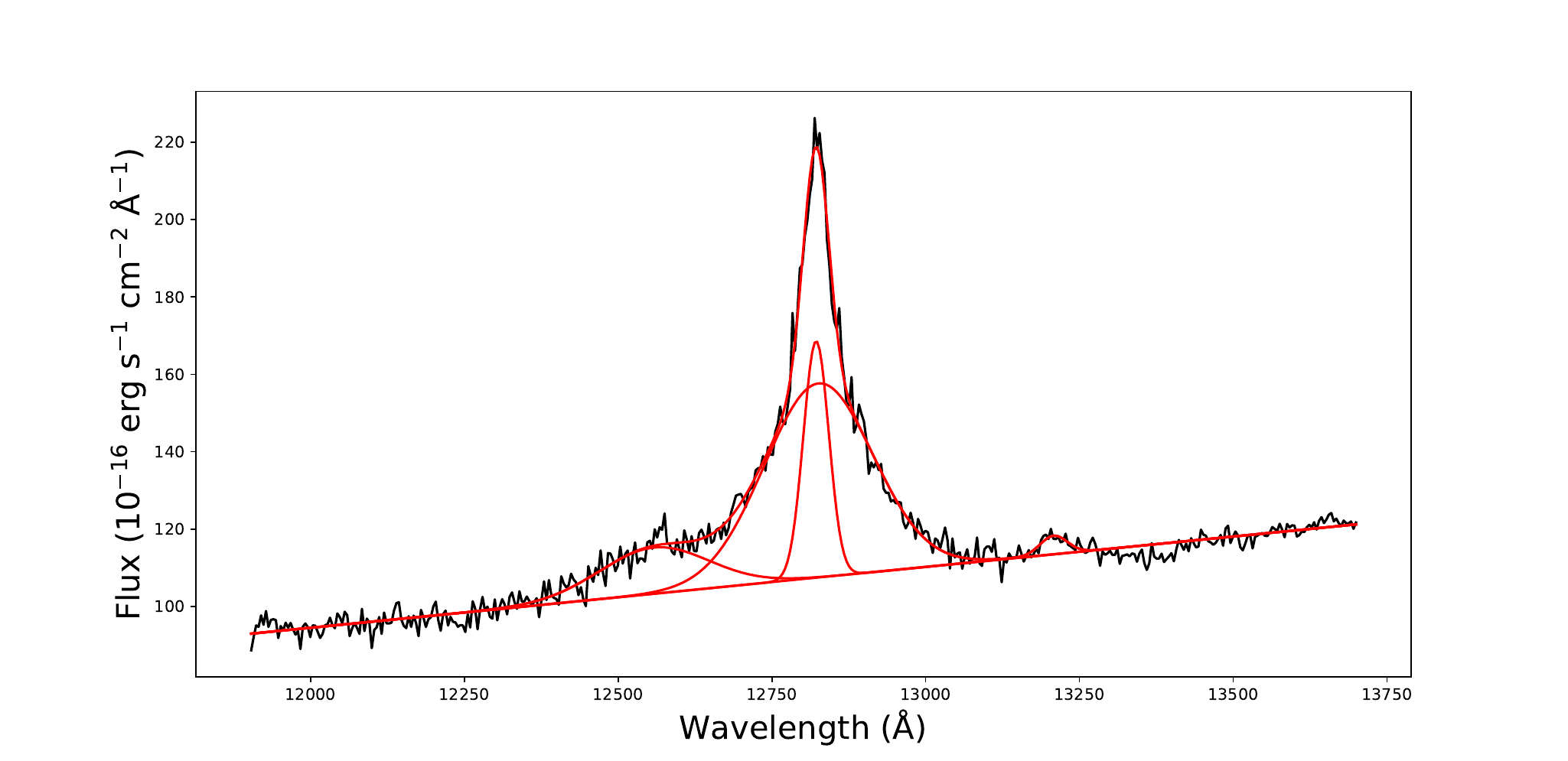}
\caption{\label{fig:line3}Composite spectrum, fit to Paschen Beta line blended with Fe II at 12570 \AA.}

\centering
\includegraphics[scale=0.5]{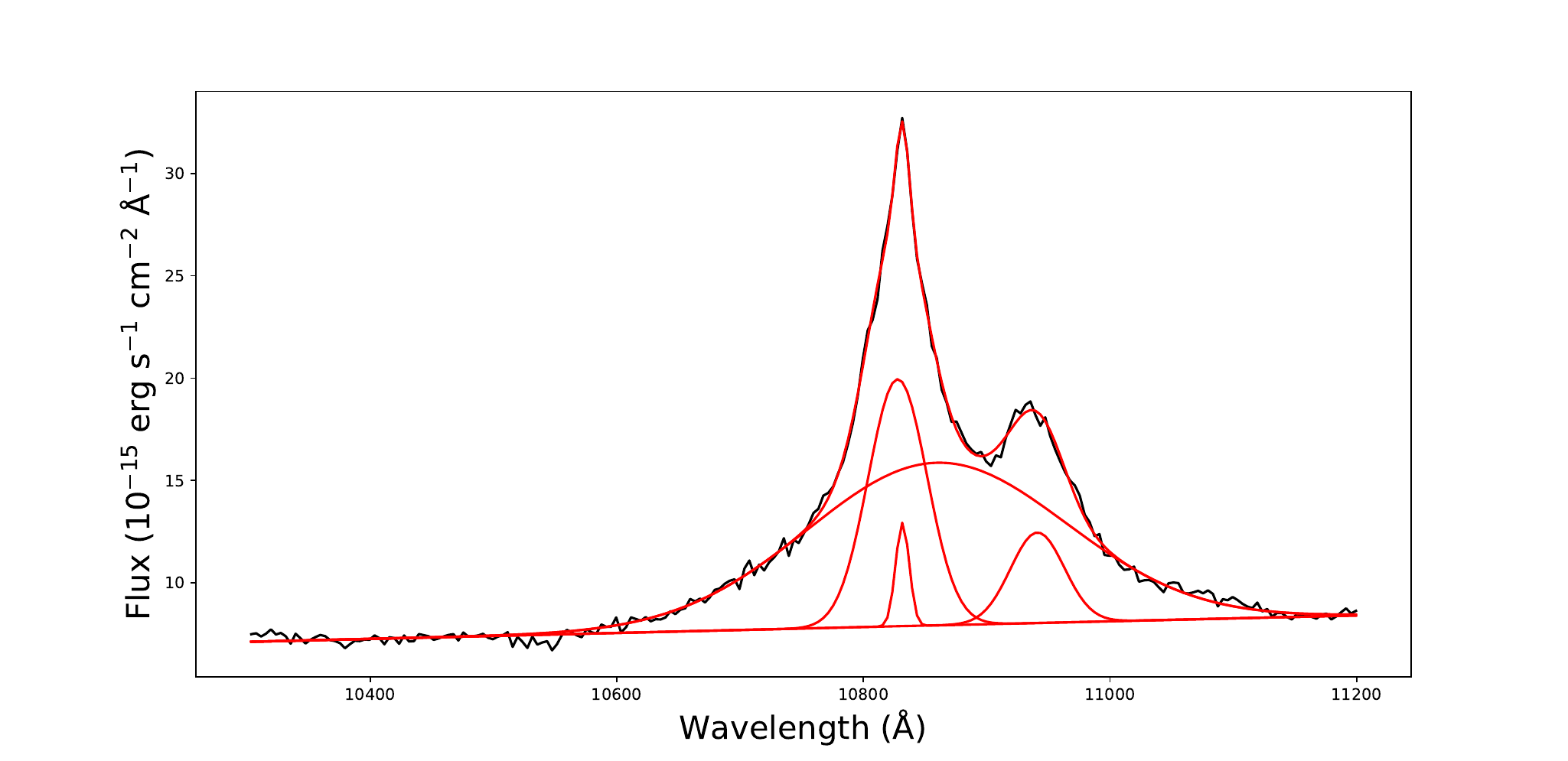}
\caption{\label{fig:line4}Composite spectrum, fit to He I 10830\AA\ blended with Paschen Gamma.}
\end{figure*}

\begin{figure*}[!ht]
\centering
\includegraphics[scale=0.5]{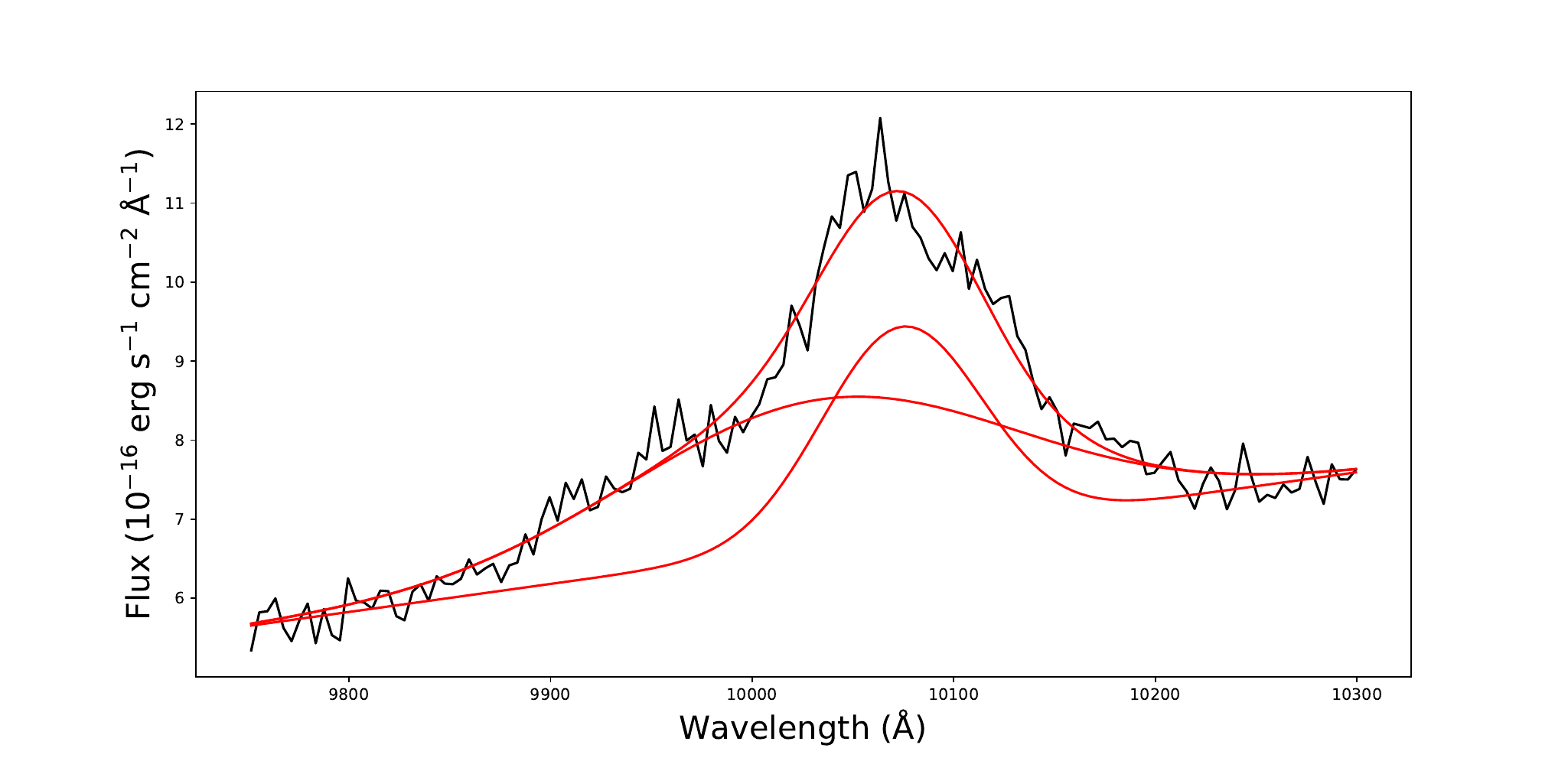}
\caption{\label{fig:line5}Composite spectrum, fit to Paschen Delta.}

\centering
\includegraphics[scale=0.5]{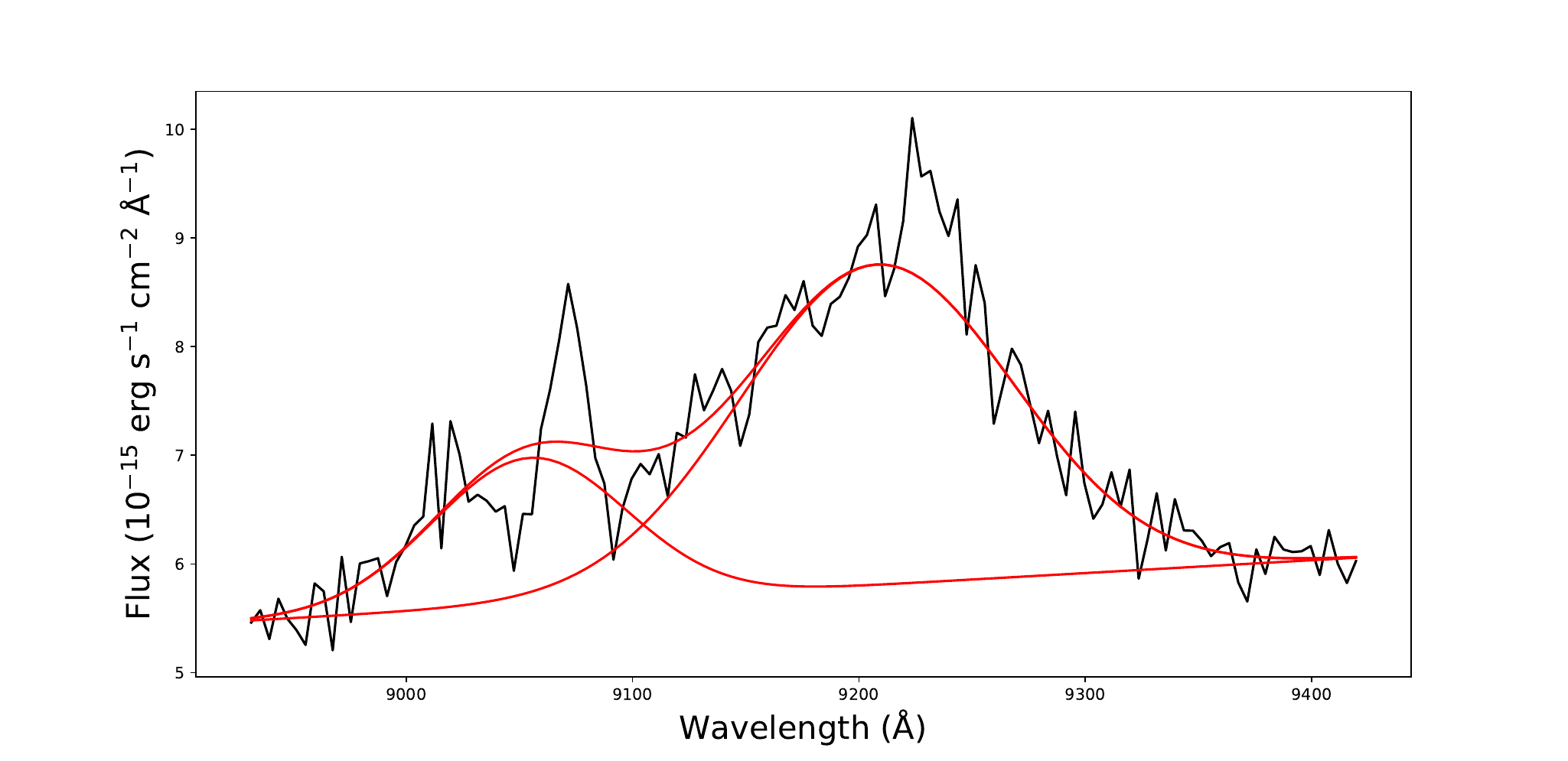}
\caption{\label{fig:line6}Composite spectrum, fit to H I (9230\AA) blended with S III 9069\AA.}
\end{figure*}

\begin{figure*}[!ht]
\centering
\includegraphics[scale=0.5]{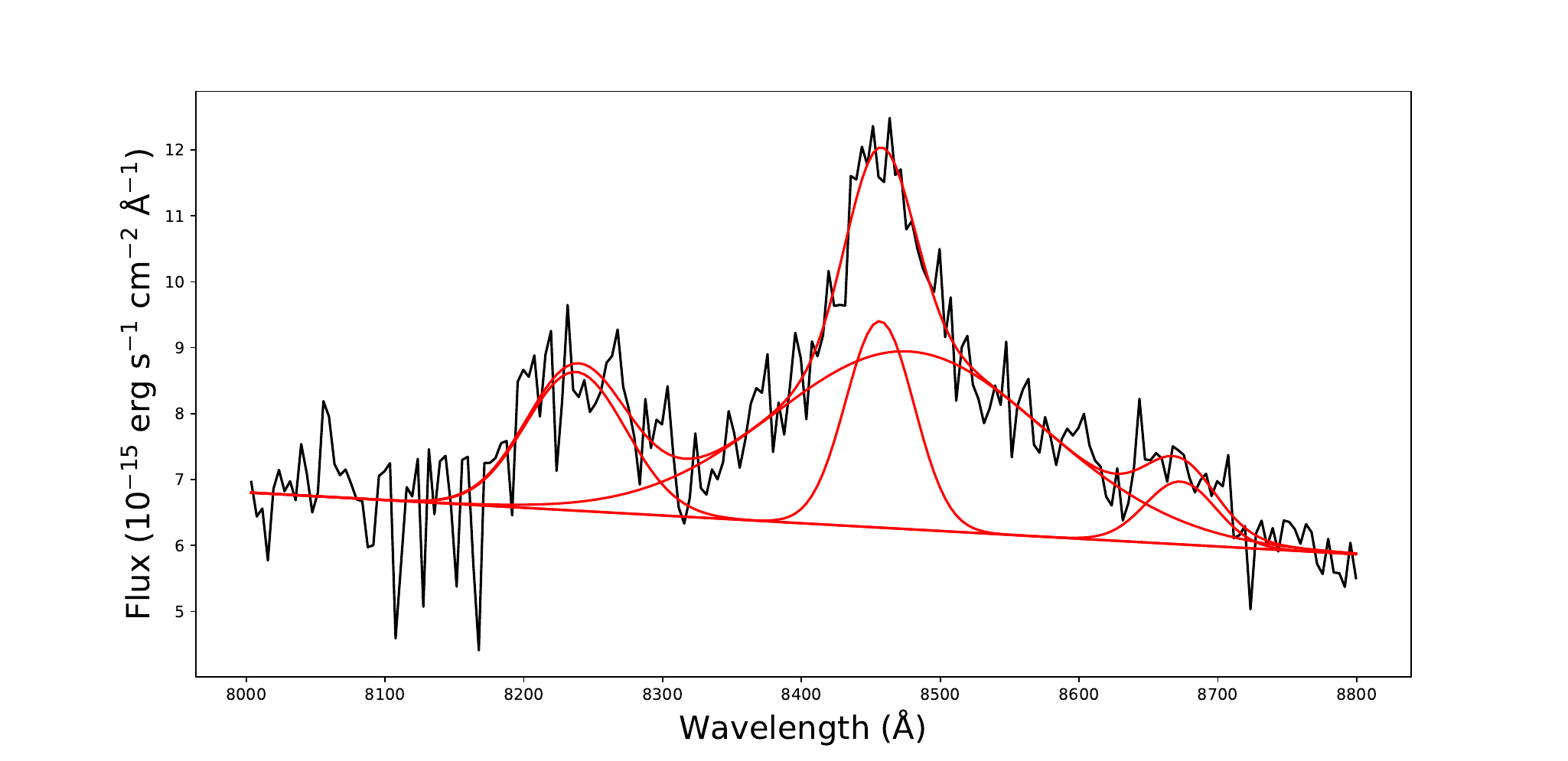}
\caption{\label{fig:line7}Composite spectrum, fit to O I (8446\AA) with other nearby weak lines deblended.}
\end{figure*}

We measured individual emission lines.  For each line or line blend, a wavelength region containing the line or line blend was defined.  In cases where deblending was necessary, Gaussians were fit to the wavelength region containing the blended lines.  The line to be measured was obtained by subtracting the Gaussian fits to the blended features.  Gaussians were also fit to unblended lines.  See Figures \ref{fig:line1}-\ref{fig:line7} for all fits.  Measurements were made of the equivalent width and integrated flux.  For the integrated flux, the line flux was integrated using Simpson's rule.  For the equivalent width, a mean was taken of the continuum flux of the line region at wavelengths greater than the upper line limit.  The normalized flux of the line was found by dividing the line flux by this mean continuum flux and subtracting this ratio from one.  The equivalent width was then the normalized flux integrated using Simpson's rule.  Table \ref{tab:Table of Measurements} provides the emission line names, central rest wavelengths from \citet{Glikman2006} and \citet{2020ApJ...888L..11S}, start and stop wavelengths for the flux integration, integrated flux, and equivalent width.  Uncertainties were found in the following way.  The standard deviation of the flux at wavelengths below the lower line limit was taken and random numbers between zero and this standard deviation were added to flux values in the emission line region.  The line measurements were done using this synthetic flux in the same way as described above.  This process was repeated 100 times and the standard deviation was taken to be the uncertainty. 

\onecolumngrid

\begin{table*}
\centering
\caption{Measurements of Emission Lines in Composite}
\label{tab:Table of Measurements}
\begin{tabular*}{\textwidth}{@{\extracolsep{\fill}} c c c c c c }
    \hline
    \hline
    {\bfseries Line} & {\bfseries Central Wavelength (\AA)} & {\bfseries Start Wavelength (\AA)} & {\bfseries Stop Wavelength (\AA)} & {\bfseries Equivalent Width (\AA)} & {\bfseries Integrated Flux (x 10$^{-12}$ erg s$^{-1}$ cm$^{-2}$)}\\
    \hline
    Brackett \textdelta & 19446 & 19350 & 19550 & 0.45 $\pm$ 0.07 & 1.346 $\pm$ 0.002\\
    Paschen \textalpha & 18750 & 18470 & 19100 & 18.9 $\pm$ 0.3 & 4.597 $\pm$ 0.002\\
    Paschen \textbeta & 12820 & 12535 & 13090 & 19.8 $\pm$ 0.4 & 1.838 $\pm$ 0.003\\
    Paschen \textgamma & 10938 & 10870 & 10995 & 6.5 $\pm$ 0.2 & 0.302 $\pm$ 0.001\\
    He I & 10830 & 10550 & 11145 & 76 $\pm$ 2 & 1.890 $\pm$ 0.003\\
    Paschen \textdelta & 10049 & 9830 & 10215 & 12.7 $\pm$ 0.9 & 0.802 $\pm$ 0.002\\
    {[}S III{]} & 9531 & 9510 & 9555 & 4.2 $\pm$ 0.2 & 0.084 $\pm$ 0.001\\
    Paschen \textsubscript{\textepsilon} & 9546 & 9420 & 9680 & 17.1 $\pm$ 0.8 & 0.467 $\pm$ 0.002\\
    H I & 9230 & 8960 & 9365 & 21 $\pm$ 1 & 0.729 $\pm$ 0.002\\
    O I & 8446 & 8280 & 8720 & 43 $\pm$ 5 & 0.901 $\pm$ 0.009\\
    \hline
\end{tabular*}
\end{table*}

\twocolumngrid

\section{Line Fits}


\end{document}